\documentclass[useAMS,usenatbib]{mn2e}

\usepackage{graphics,graphicx}
\usepackage{epstopdf}
\usepackage{epsfig}

\title[SN~2013df]{The Continuing Story of SN IIb 2013df: New Optical and IR Observations and Analysis}

\author[Szalai et al.]{Tam\'{a}s Szalai$^{1}$, J\'{o}zsef Vink\'{o}$^{1,2}$, Andrea P. Nagy$^{1}$, Jeffrey M. Silverman$^{2,3}$, 
\newauthor J. Craig Wheeler$^{2}$, Govinda Dhungana$^{4}$, G. Howie Marion$^{2}$, Robert Kehoe$^{4}$,
\newauthor Ori D. Fox$^{5,6}$, Kriszti\'{a}n S\'{a}rneczky$^{7,8}$, G\'{a}bor Marschalk\'{o}$^{7,9}$, Barna I. B\'{i}r\'{o}$^{9}$,
\newauthor Tam\'{a}s Borkovits$^{8,9}$, Tibor Heged\"{u}s$^{9}$, R\'{o}bert Szak\'{a}ts$^{7}$, Farley V. Ferrante$^{4}$,
\newauthor Evelin B\'{a}nyai$^{7,10}$, Gabriella Hodos\'{a}n$^{7,11}$, J\'{a}nos Kelemen$^{7}$, Andr\'{a}s P\'{a}l$^{7}$\\
\\
$^{1}$Department of Optics and Quantum Electronics, University of Szeged, H-6720 Szeged, D\'om t\'er 9., Hungary\\
$^{2}$Department of Astronomy, University of Texas at Austin, Austin, TX 78712-1205, USA\\
$^{3}$NSF Astronomy and Astrophysics Postdoctoral Fellow\\
$^{4}$Department of Physics, Southern Methodist University, Dallas, TX 75275, USA\\
$^{5}$Department of Astronomy, University of California, Berkeley, CA 94720-3411, USA\\
$^{6}$Space Telescope Science Institute, 3700 San Martin Drive, Baltimore, MD 21218, USA\\
$^{7}$Konkoly Observatory, Research Centre for Astronomy and Earth Sciences, Hungarian Academy of Sciences,\\
\hspace{0.4mm} Konkoly Thege Mikl\'os \'ut 15-17, H-1121 Budapest, Hungary\\
$^{8}$ELTE Gothard-Lend\"{u}let Research Group, H-9700 Szombathely, Szent Imre herceg \'{u}t 112, Hungary\\
$^{9}$Baja Astronomical Observatory of University of Szeged, H-6500 Baja, Szegedi \'{u}t, Kt. 766, Hungary\\
$^{10}$Department of Physics of Complex Systems, Lor\'and E\"otv\"os University, P\'azm\'any P. s\'et\'any 1/A, H-1117 Budapest, Hungary\\
$^{11}$School of Physics \& Astronomy, University of St Andrews, St Andrews, KY16 9SS, UK}

\begin{document}

\date{in original form 2015}

\pagerange{\pageref{firstpage}--\pageref{lastpage}} \pubyear{2015}

\maketitle

\label{firstpage}

\begin{abstract}

SN~2013df is a nearby Type IIb supernova that seems to be the spectroscopic twin of the well-known SN~1993J.
Previous studies revealed many, but not all interesting properties of this event. Our goal was to add new understanding of both the early and late-time phases of SN~2013df.
Our spectral analysis is based on 6 optical spectra obtained with the 9.2m Hobby-Eberly Telescope during the first month after explosion, complemented by a near-infrared spectrum. 
We applied the SYNAPPS spectral synthesis code to constrain the chemical composition and physical properties of the ejecta. 
A principal result is the identification of ``high-velocity'' He\,\begin{small}I\end{small} lines in the early spectra of SN~2013df, manifest as the blue component of the double-troughed profile at $\sim$5650 \AA. 
This finding, together with the lack of clear separation of H and He lines in velocity space, indicates that both H and He features form at the outer envelope during the early phases.
We also obtained ground-based ${\it BVRI}$ and ${\it g'r'i'z'}$ photometric data up to +45 days and unfiltered measurements with the ROTSE-IIIb telescope up to +168 days.
From the modelling of the early-time quasi-bolometric light curve, we find $M_\rmn{ej} \sim$ 3.2$-$4.6 $M_\rmn{\odot}$ and $E_\rmn{kin} \sim$ 2.6$-$2.8 $\times$ 10$^{51}$ erg for the 
initial ejecta mass and the initial kinetic energy, respectively, which agree well with the values derived from the separate modelling of the light-curve tail.
Late-time mid-infrared excess indicates circumstellar interaction starting $\sim$1 year after explosion, in accordance with previously published optical, X-ray, 
and radio data.

\end{abstract}


\section{Introduction}\label{intro}

Type IIb supernovae (SN~IIb) are thought to arise from the core-collapse of massive ($M >$ 8 $M_\rmn{\odot}$) stars that lost most, but not all, of their thick H envelopes 
prior to explosion. This explains the special spectral evolution characteristic of the IIb subtype. The most conspicuous features are H Balmer lines that are
strong around maximum light but weaken relatively quickly after that and He lines that become strong after the maximum. This makes SN IIb members of an intermediate group between Type II and
the hydrogen-poor, stripped envelope Type Ib/c explosions.
The first identified case of the subtype was SN~1987K \citep{Filippenko88}, while the best known member of the group is one of the closest and brightest SN of recent 
decades, SN~1993J \citep[see e.g.][]{Filippenko93,Wheeler93,Richmond94,Barbon95,Matheson00}.

While SN~IIb constitute approximately 10-12\% of all core-collapse SNe \citep{Li11}, detailed analyses have been published only on about a dozen of them.
As from every type of supernovae, one of the main questions is the nature of the progenitor stars. Direct identification 
of the progenitor has been possible only in four cases: SN~1993J \citep[a K-type supergiant with initial mass $M_\rmn{in} \sim$13-22 $M_\rmn{\odot}$, which may be a component of 
an interacting binary system, see e.g.][]{VanDyk02,Maund04,Maund09,Fox14}, 
SN~2008ax \citep[a Wolf-Rayet star with $M_\rmn{in} \sim$10-28 $M_\rmn{\odot}$, probably exploded in a binary system,][]{Crockett08,Pastorello08,Folatelli15},
SN~2011dh \citep[a yellow supergiant with $M_\rmn{in} \sim$12-15 $M_\rmn{\odot}$,][]{Maund11,VanDyk11,VanDyk13,Ergon14}, and SN~2013df (see details later).
In other cases, analysis of early light curves and/or the interactions with the circumstellar matter (CSM) originating from pre-explosion mass-loss processes offer a chance to constrain the exploding star.
Spectral analysis of light echoes showed that Cassiopeia A (hereafter Cas A) was also a Type IIb explosion \citep{Krause08}. The detailed study of this object can provide valuable pieces of information about 
the explosion mechanism of SN IIb, which are difficult to extract from studying of the early supernova phases. Such an interesting result was the confirmation of the asymmetric explosion of Cas A 
\citep{DeLaney10,Rest11,Fesen16}.

Based on the observational characteristics, SN IIb may be divided into two subgroups \citep{Chevalier10}. SN~1993J and some similar explosions, e.g. SNe~2011hs 
\citep{Bufano14} and 2013fu \citep{Kumar13,MG15}, are thought have arisen from massive stars with extended H envelopes ($R \sim$ 10$^{13}$ cm or some hundreds of solar radii). 
Other SN~IIb seem to have much more compact ($R \sim$ 10$^{11}$ cm, $\sim$ 1-2 $R_{\odot}$) progenitors, e.g. SN~2008ax, 
SN~2003bg or SN~2001ig. Members of these two subgroups are sometimes called SN~eIIb (extended IIb) and SN~cIIb (compact IIb), respectively.
Nevertheless, the classification of SN~2011dh was ambiguous at first. While it was classified as a cIIb based on its early-time luminosity and radio emission \citep{Arcavi11}, 
later studies -- based on the modeling of the bolometric light curve \citep{Bersten12}, on comparative multiwavelength studies \citep{Horesh13}, and on study of late-time X-ray data \citep{Maeda14} -- 
strengthen the interpretation as an intermediate case ($R_{in} \sim$ a few tens of $R_{\odot}$) in agreement with the results of the direct progenitor identification (see above).
This case illustrates that there are several open questions concerning this sub-classification. Moreover, there are other aspects which should be taken into account, e.g. the 
interpretation of early-time radio data \citep[see e.g.][]{Bufano14} or the possible asphericity of the explosions \citep{Mauerhan15}.
Nonetheless, as \citet{BenAmi15} report, the amount of the early-time UV excess can be also a good indicator of the size of the progenitor 
(although, as the authors also noted, the number of the studied SNe is too low to draw a very general conclusion).
The reality is likely to be complicated, because, based on theoretical and observational results, the progenitors of SN IIb explosions may be members of interacting binary systems 
\citep[see e.g.][]{Woosley94,Maund04,Maund07,Silverman09,Dessart11,Claeys11,Benvenuto13}.

A well-known general characteristic of SN IIb is the presence of double-peaked light curves in the whole optical range, which was first observed in the case of SN~1993J. 
The rapid decline after the initial peak is interpreted as adiabatic cooling of the ``fireball'' after the SN shock has broken out through the star's surface; the
timescale of the cooling depends mainly on the radius of the progenitor \citep[see e.g.][]{Chevalier08,Bersten12}. 
As a result of thermalization of $\gamma$-rays and positrons originating from the radioactive decay of $^{56}$Ni and $^{56}$Co, the light
curves reach a secondary maximum. The length of the initial declining phase is usually some days, so the fireball phase is not easy to detect (see Section \ref{lc_lc}).
Up to now, detailed analyses have been published on only a few SN IIb observed from the very early phases. Thus, it is important to study well-observed individual objects as thoroughly as possible.

SN~2013df is a recently discovered, nearby SN~IIb.
This object, located 32$''$ east and 14$''$ south from the center of the spiral galaxy NGC~4414, was discovered on June 8, 2013 \citep{Ciabattari13}.
The first spectrum, obtained by \citet{Cenko13},
showed clear resemblance to the early spectra of SN~1993J, which suggested that SN~2013df was a SN IIb. This suspicion was verified by 
\citet{VanDyk14} (hereafter VD14), who presented early photometric and spectroscopic data and reached conclusions concerning the properties of the progenitor, 
identifying it as a yellow supergiant star with an estimated initial mass of $M_\rmn{in} \sim$13-17 M$_{\odot}$, an effective temperature of $T_\rmn{eff} \sim$4250 K, and an initial radius of 
$R_\rmn{eff} \sim$545 $R_{\odot}$. \citet{MG14} (hereafter MG14) presented a detailed analysis of early UV-optical light curves and some spectra, including nebular ones. 
Their results concerning the progenitor are compatible with those of VD14 ($M_\rmn{in} \sim$12-13 M$_{\odot}$, $R_\rmn{eff} >$ 64$-$169 $R_{\odot}$). The common main conclusion 
of these authors is that SN~2013df is very similar to SN~1993J, and it was very probably the endpoint of a massive star with an extended but not massive H envelope.
The similarity between SNe 1993J and 2013df was strengthened by the work of \citet{BenAmi15} (hereafter BA15) based on the comparison of early UV-optical spectra 
of several SN IIb. 
Results based on late-time optical spectroscopy \citep{Maeda15}, as well as on early and late-time radio and X-ray observations \citep{Kamble16}, suggest the presence of 
CSM interactions. This evidence indicates that the extended progenitors, as for SN~2013df, suffer from substantial mass loss some years before the explosion, which may be caused by 
the assumed binary companions of the exploding stars.

In this paper, we present some new results concerning the early and late-time properties of SN~2013df. First, we describe our ground-based spectroscopic and photometric 
observations in Section \ref{obs}. The method and results of modelling of spectra using a spectral synthesis code are shown in Section \ref{spec}.
In Section \ref{lc}, we present the steps of our analysis of the photometric data, including the comparative analysis of early light curves 
with those of other SN~IIb, and the extraction of explosion parameters from early-time bolometric light curve modelling and from late-time light curve analysis.
At the end of Section \ref{lc}, we also present our findings concerning the analysis of late-time mid-infrared data of SN~2013df.
Finally, in Section \ref{conc}, we discuss our results and present our conclusions.

\section{Observations and data reduction}\label{obs}

During the reduction and analysis of our data, we used the adopted values of important parameters of SN~2013df and of its host galaxy, NGC~4414.
We adopted the explosion date $t_0$ = 2,456,447.8 $\pm$ 0.5 JD, or June 4.3 UT (determined by VD14 from the overall comparison of the SN 2013df light 
curves with those of SNe 1993J and 2011dh) to calculate the epochs of both spectroscopic and photometric measurements.

We used the distance modulus of the host galaxy $\mu_0$= 31.10 $\pm$ 0.05 mag ($D$ = 16.6 $\pm$ 0.4 Mpc), established by 
\citet{Freedman01} based on their study of Cepheid stars. This value was used also by VD14. This distance modulus is definitely lower than the one used by MG14 
($\mu_0$= 31.65 $\pm$ 0.30 mag), which results in significant differences in the calculated bolometric magnitudes and luminosities of the SN (see Section \ref{lc}).
(Note that the value used by MG14 cannot be the weighted mean value of distance moduli provided by the NASA/IPAC extragalactic database, 
NED\footnote{http://ned.ipac.caltech.edu.}, as the authors refer to it; the mean value of individually referenced moduli listed in NED is $\mu_0$= 31.23 $\pm$ 0.54 mag).
The lower value of the distance is also suggested by the published values of the redshift of the host, $z$ = 0.002388 \citep[given by NED,][]{Rhee96}, and $z$ = 0.002874 
(determined from the position of Na \begin{small}I\end{small} D lines associated with the host, VD14), which correspond to distances of $\sim$10.3 Mpc and $\sim$12.4 Mpc, 
respectively \citep{Wright06}.

We adopted $E(B-V)$ = 0.09 $\pm$ 0.01 mag as the total reddening for SN 2013df from the work of VD14 who determined this value via analyzing the Galactic and host-galaxy 
components of Na\,\begin{small}I\end{small} D in their high-resolution spectra.

The adopted values are also shown in Table \ref{tab:ad_par}.

\begin{table}
\begin{center}
\caption{Adopted values that are used throughout the paper}
\label{tab:ad_par}
\begin{tabular}{ccc}
\hline
\hline
Parameter & Adopted value & Reference \\
\hline
$t_0$ & 2,456,447.8 $\pm$ 0.5 JD & 1 \\
$\mu_0$ & 31.10 $\pm$ 0.05 mag & 2 \\
$E(B-V)_{total}$ & 0.09 $\pm$ 0.01 mag & 1 \\
\hline
\end{tabular}
\end{center}
\smallskip
Notes. References: (1) \citet{VanDyk14}; (2) \citet{Freedman01}.
\end{table}

\subsection{Spectroscopy}\label{obs_sp}

\begin{figure*}
\begin{center}
\leavevmode
\includegraphics[width=.6\textwidth]{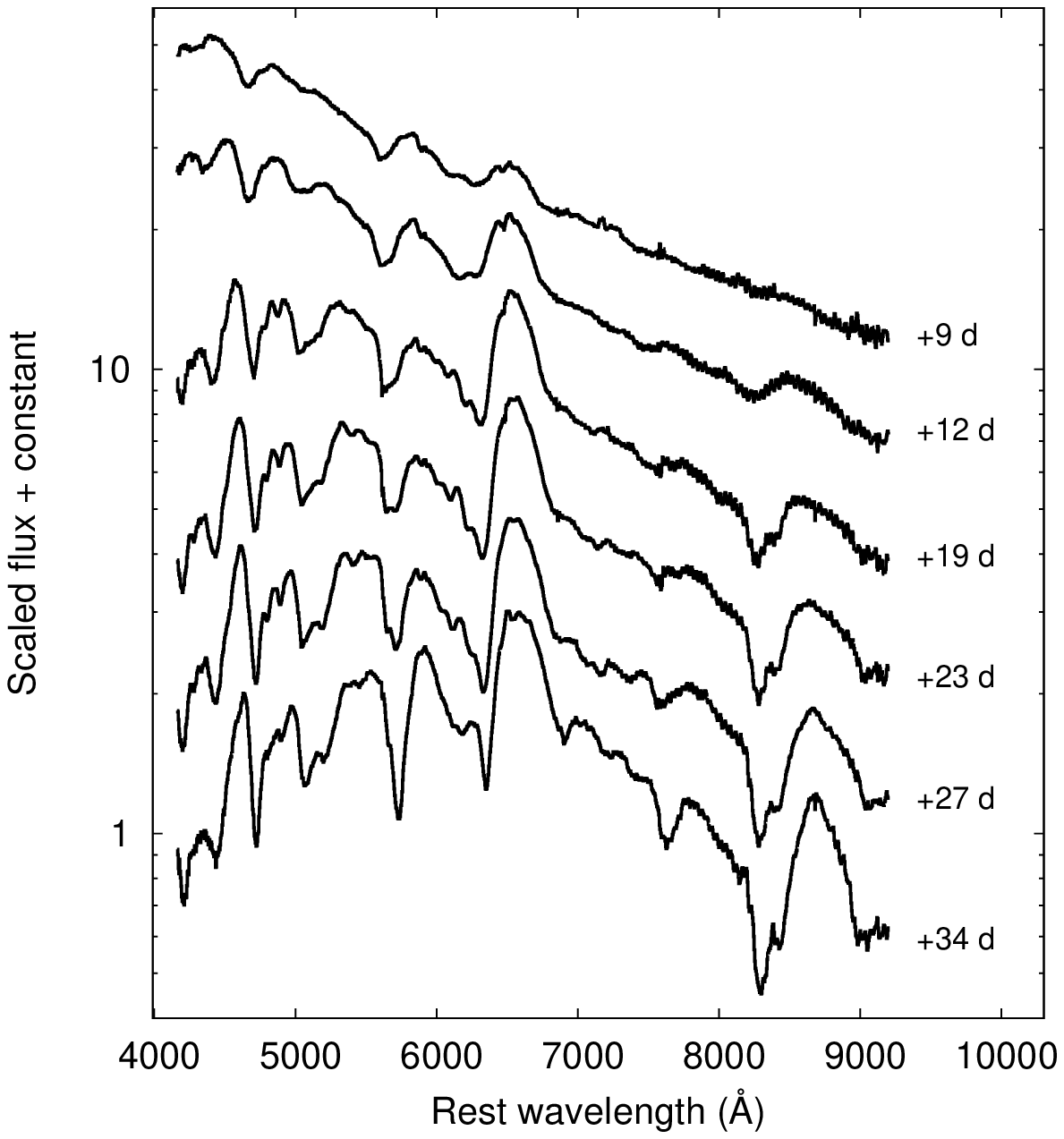} \hspace{5mm}
\includegraphics[width=.6\textwidth]{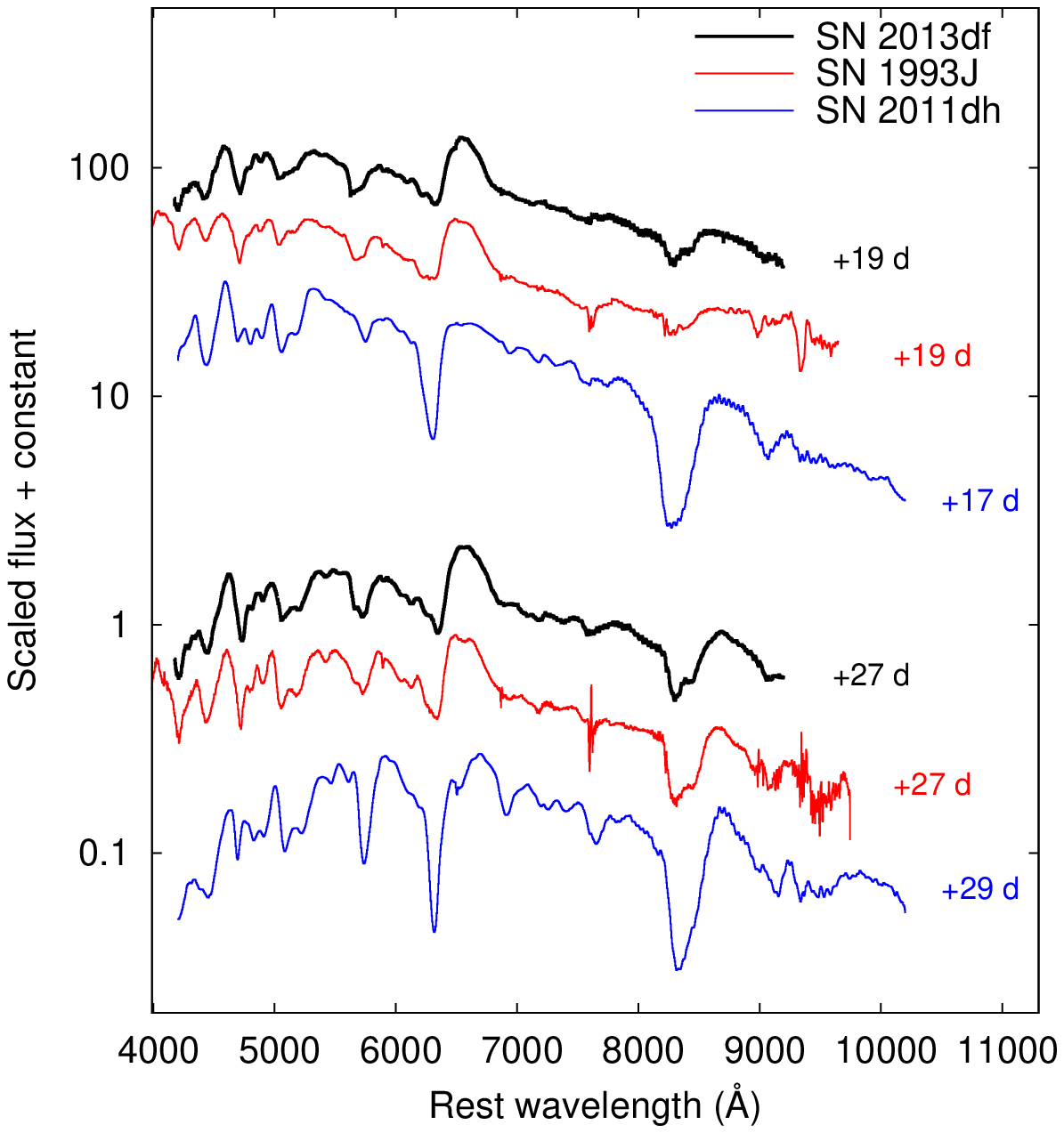}
\end{center}
\caption{{\it Top:} Observed HET spectra of SN~2013df. Phases are given relative to the explosion date
($t_{0}$ = 2,456,447.8 $\pm$ 0.5 JD, or June 4.3 UT) determined by \citet{VanDyk14}.
{\it Bottom:} Spectra of SN~2013df compared with spectra of other SN IIb 1993J \citep{Barbon95} and 2011dh \citep{Marion14} at similar ages.}
\label{fig:sp}
\end{figure*}

We collected a sample of high-quality spectroscopic data on SN~2013df. The object was monitored with the Marcario Low Resolution Spectrograph (LRS)
on the 9.2m Hobby-Eberly Telescope (HET) at McDonald Observatory. Table \ref{tab:spec} summarizes the basic details of the 6 spectra (R = 300) taken
between June 13 and July 8, 2013, during the first month after explosion. 
All of our HET spectra were reduced using standard techniques \citep[e.g.][]{Silverman12}. Routine CCD processing and spectrum extraction were completed with 
IRAF\footnote{IRAF is distributed by the National Optical Astronomy Observatories, which are operated by the Association of Universities for 
Research in Astronomy, Inc., under cooperative agreement with the National Science Foundation.}. We obtained the wavelength scale from low-order
polynomial fits to calibration-lamp spectra. Small wavelength shifts were then applied to the data after cross-correlating a template sky to the
night-sky lines that were extracted with the SN. Using our own IDL routines, we fit spectrophotometric standard-star spectra to the data in
order to flux calibrate our spectra and to remove telluric lines \citep[see e.g.][]{Wade88,Matheson00}.
The top panel of Fig. \ref{fig:sp} shows the observed HET spectra of SN~2013df, while the comparison of 
these data with spectra of SNe 1993J and 2011dh is shown in the bottom panel.

\begin{table}
\begin{center}
\caption{Log of spectral observations obtained with HET LRS and IRTF.}
\label{tab:spec}
\begin{tabular}{ccccc}
\hline
\hline
UT Date & Phase & Instrument & Range & R \\
~ & (days) & ~ & (\AA) & ($\lambda$/$\Delta\lambda$) \\
\hline
2013-06-13 & +9 & HET LRS & 4172-10,800 & 300 \\
2013-06-16 & +12 & HET LRS & 4172-10,800 & 300 \\
2013-06-23 & +19 & HET LRS & 4172-10,800 & 300 \\
2013-06-27 & +23 & HET LRS & 4172-10,800 & 300 \\
2013-07-01 & +27 & HET LRS & 4172-10,800 & 300 \\
2013-07-02 & +28 & IRTF & 6500-25,430 & 200 \\
2013-07-08 & +34 & HET LRS & 4172-10,800 & 300 \\
\hline
\end{tabular}
\end{center}
\smallskip
{\bf Notes.} Phases are given relative to the explosion date ($t_0$ = 2,456,447.8 $\pm$ 0.5 JD, or June 4.3 UT) determined by \citet{VanDyk14}.
\end{table}

Additionally, a low resolution (R$\approx$200, $\lambda$=0.65--2.5 $\mu$m) near-infrared (NIR) spectrum was also obtained on July 2 (+28 days) with the 
3 meter telescope at the NASA Infrared Telescope Facility (IRTF) using the SpeX medium-resolution spectrograph \citep{Rayner03}. IRTF data are reduced using a 
package of IDL routines specifically designed for the reduction of SpeX data \citep[Spextool v. 3.4;][]{Cushing04}.

\subsection{Photometry}\label{obs_phot}

The photometric observations for SN~2013df were obtained using three different ground-based telescopes. We used the 0.6/0.9m Schmidt-telescope (Piszk\'{e}stet\H{o} 
Mountain Station of Konkoly Observatory, Hungary) with the attached 4096 $\times$ 4096 CCD (FoV 70 $\times$ 70 arcmin$^2$) equipped with Bessel {\it BVRI} filters; 
the 0.5m Baja Astronomical Robotic Telescope (BART, Baja Observatory, Hungary) with a 4096 $\times$ 4096 front illuminated Apogee U16 CCD (FoV 40 $\times$ 40 arcmin$^2$; 
the frames were taken with 2 $\times$ 2 binning) equipped with Sloan {\it g'r'i'z'} filters; and the 0.45m ROTSE-IIIb telescope at McDonald Observatory, 
operated with an unfiltered CCD with broad wavelength transmission from 3,000 \AA\ to 10,600 \AA.

Photometric follow-up observations with the first two telescopes started at +8d, and continued up to +45d. 
To obtain the Konkoly {\it BVRI} magnitudes, we carried out PSF-photometry on the SN and two local comparison (tertiary standard) stars using the {\it daophot/allstar} task in 
IRAF. We applied an aperture radius of 6\arcsec and a background annulus from 7\arcsec to 12\arcsec for SN~2013df as well as for the local comparison stars.
The instrumental magnitudes were transformed to the standard system applying the following equations:

\begin{eqnarray}
V - v = C_\rmn{V} \cdot (V-I) + \zeta_\rmn{V} \nonumber\\
(B-V) = C_\rmn{BV} \cdot (b-v) + \zeta_\rmn{BV} \nonumber\\
(V-R) = C_\rmn{VR} \cdot (v-r) + \zeta_\rmn{VR} \nonumber\\
(V-I) = C_\rmn{VI} \cdot (v-i) + \zeta_\rmn{VI},
\label{eq:standtr}
\end{eqnarray}

\noindent where lowercase and uppercase letters denote instrumental and standard magnitudes, respectively.
The color terms ($C_\rmn{X}$) were determined by measuring Landolt standard stars in the field of PG1633 \citep{Landolt92} observed during photometric conditions:
$C_\rmn{V}$ = $-$0.025, 
$C_\rmn{BV}$ = 1.268, 
$C_\rmn{VR}$ = 1.024, 
$C_\rmn{VI}$ = 0.964. 
These values were kept fixed while computing the standard magnitudes for the whole dataset.
Zero-points  ($\zeta_\rmn{X}$) for  each  night  were  measured using the local tertiary standard stars mentioned above. These  local  comparison
stars were tied to the Landolt standards during the photometric calibration.

The {\it g'r'i'z'} data from Baja Observatory were standardized using $\sim$100 stars within the $\sim$40 $\times$ 40 arcmin$^2$ field-of-view around the SN, 
taken from the SDSS DR12 catalog. In order to avoid selecting saturated stars from the SDSS catalog, a magnitude cut 14 $<$ {\it r'} $<$ 18 was applied during 
the photometric calibration.

The SN 2013df position is outside the regular ROTSE-III supernova search fields; however, we scheduled the ROTSE-IIIb
telescope for extended follow-up. Up to 30 exposures were taken per night of observation when the weather was
supportive. Open CCD data were taken from +11d to +82d, and then from +157d to +237d. In all cases, data from each
day are co-added. The ROTSE data were first calibrated to USNO B1.0 R2-band. A deep template of the host was taken in
early 2015 when the SN was well below the detection limit. This template was used to assess the sky-subtracted host
flux in the photometric aperture for each epoch, accounting for focus variations during the observing period.
Because of the position of the source near the host nucleus, we compared the co-adds of the first and second half of the
nightly images to extract a systematic uncertainty per epoch. We also examined the photometry of nearby reference
stars and extracted a separate systematic uncertainty from the rms of reference star residuals at each epoch.

During the reduction of ROTSE data, epochs with rms $>$0.4 mag were rejected from the light curve. All systematic uncertainties are considered to be uncorrelated
epoch-to-epoch. The ROTSE data are compared to interpolated Konkoly R data from +9d to +40d, and an offset
correction of 0.030 magnitude is obtained with a $~\chi{^2}/dof=1.04$.
We took the rms value ($0.175$ magnitude) of the residuals around this fit as an additional uncorrelated systematic uncertainty on these points. 
We utilized the R-band data for our measurements and derived an offset correction of this data relative to the Konkoly R-band data from +7d to +43d. 
This yields a 0.05 magnitude correction to MG14, with a $~\chi{^2}/dof = 1.93$. The data exhibit some deviation from the corresponding 
Konkoly data early and late in the second peak. 

\begin{figure}
\begin{center}
\leavevmode
\includegraphics[width=.5\textwidth]{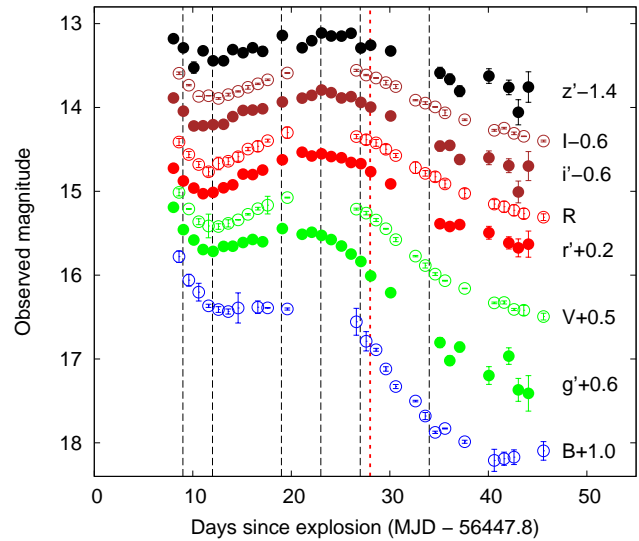}
\end{center}
\caption{Standard optical light curves of SN 2013df (open circles: {\it BVRI}, filled circles: {\it g'r'i'z'}). The vertical dashed lines mark the epochs of spectral observations 
(dashed lines: HET LRS, red dotted line: IRTF).}
\label{fig:lc}
\end{figure}

The Konkoly {\it BVRI},  Baja {\it g'r'i'z'}, and unfiltered ROTSE-IIIb magnitudes are presented in Tables \ref{tab:mag1}, \ref{tab:mag2}, and \ref{tab:rotse_lc}, respectively. 
Errors (given in parentheses) contain both the uncertainties of the photometry and the standard transformation (or, in the case of ROTSE magnitudes, the uncertainties of calibrations).
The standard {\it BVRI} and {\it g'r'i'z'} light curves are shown in Figure \ref{fig:lc}, in which we also marked the epochs of our spectral observations.
The ROTSE data, as well as the comparison of different light curves and their detailed analysis are presented in Section \ref{lc}.

The optical data were supplemented by the available {\it Swift/UVOT} data. We reduced these data using standard HEAsoft tasks: individual frames were 
summed with the {\it uvotimsum} task, and magnitudes were determined via aperture photometry using the task {\it uvotsource}. Since our results agree 
within the uncertainties with the ones published by MG14, we do not analyze these data in detail. At the same time, we used the Swift magnitudes to construct 
the bolometric light curve (see Section \ref{lc_bol}).

\section{Spectral analysis}\label{spec}

While VD14 published only one
spectrum obtained at +37 days, MG14 presented a complex study partly based on four photospheric and two nebular spectra. In both papers, the line 
identification was based on spectral comparison of SN~2013df and other, previously studied SN IIb.
BA15 presented four HST UV-optical spectra obtained in the photospheric phase. They used their own Monte Carlo
radiative-transfer code in order to produce synthetic spectra and study the structure 
of the ejecta of SN~2013df. While they got a good match between the observed and synthetic spectra at wavelengths longer than 5000 \AA, their fits are
less satisfying at shorter wavelengths (as the authors noted, the reason for the underestimation of the fluxes below 5000 \AA\ might be 
the interaction of the SN blast wave and circumstellar material).

While the cited papers contain many interesting results concerning SN~2013df, our conclusions based on 
the thorough analysis of our well-sampled spectral dataset obtained over the first month after explosion can deepen this understanding. In this section, we describe the results 
of our modelling carried out using a spectrum synthesis code, and present our findings concerning the evolution of the main spectral components, 
especially H\,\begin{small}I\end{small} and He\,\begin{small}I\end{small} lines.

\subsection{Spectral modelling}\label{sp_mod}

We applied the parameterized resonance scattering code SYN++ in combination with SYNAPPS\footnote{Software was retrieved from https://c3.lbl.gov/es/} 
\citep{Thomas11} to model the spectroscopic evolution of SN~2013df (hereafter we refer only to SYNAPPS). 
Detailed spectral modelling based on this code -- or on SYNOW, the original version of these kind of codes \citep[see e.g.][]{Jeffery90,Branch01} -- were 
carried out only in a few cases of SN~IIb: 1993J \citep{Elmhamdi06}, 2010as \citep{Folatelli14}, 2011dh \citep{Sahu13,Marion14}, 2011ei 
\citep{Milisavljevic13}, and 2011fu \citep{Kumar13}. Since the spectral similarity is very close between SNe 2013df and 1993J (see the bottom panel 
of Figure \ref{fig:sp}), we used \citet{Elmhamdi06} as a starting point in our modelling, as well as the paper of \citet{Mazzali09} about
SN~2003bg.
Before the fitting, all spectra were corrected for the total reddening and the redshift of the host galaxy. 

\begin{figure*}
\begin{center}
\includegraphics[width=.45\textwidth]{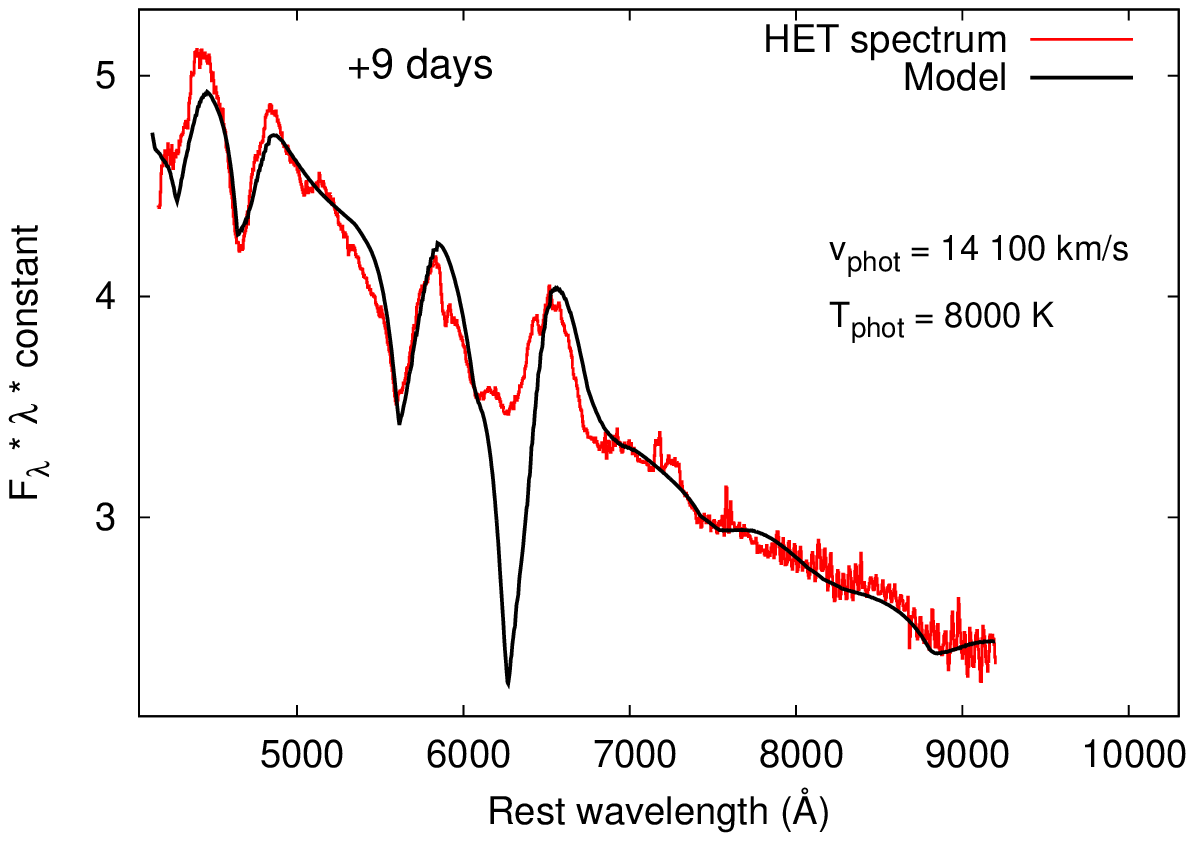} \hspace{5mm}
\includegraphics[width=.45\textwidth]{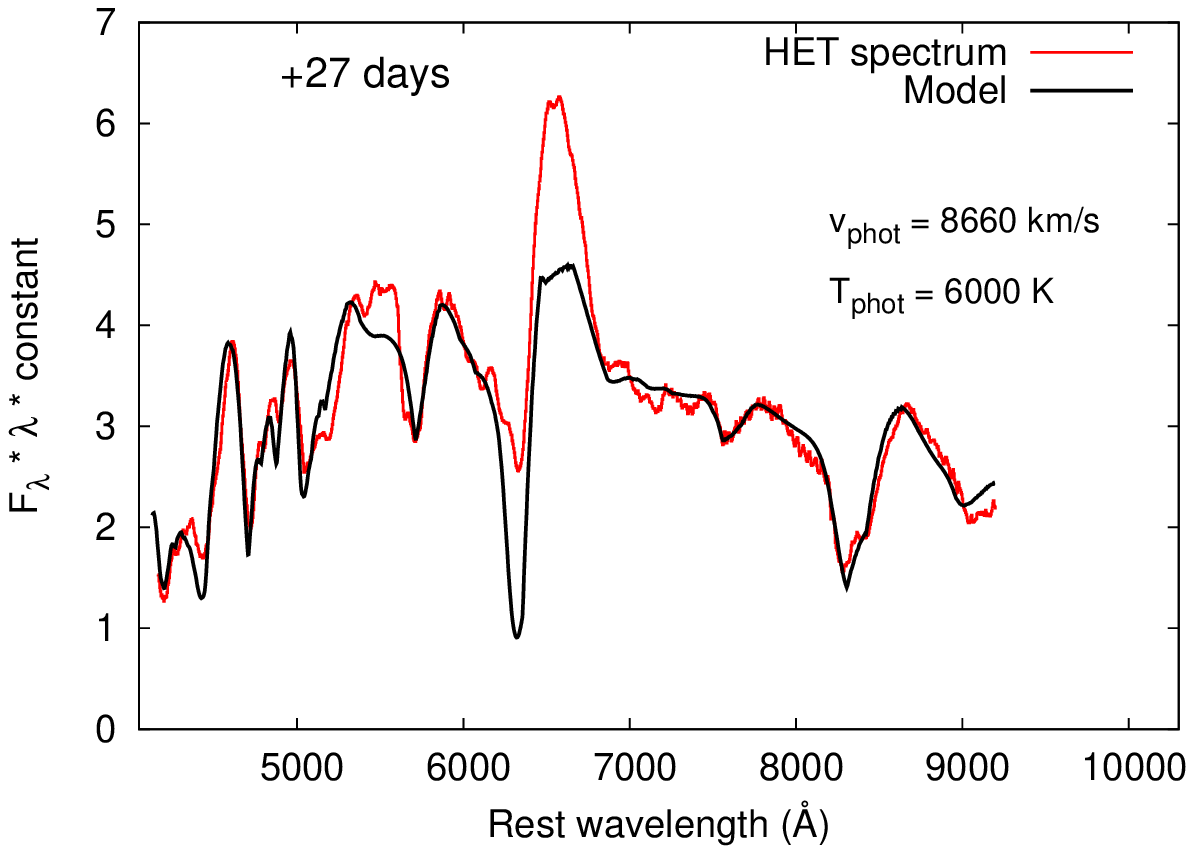}
\includegraphics[width=.45\textwidth]{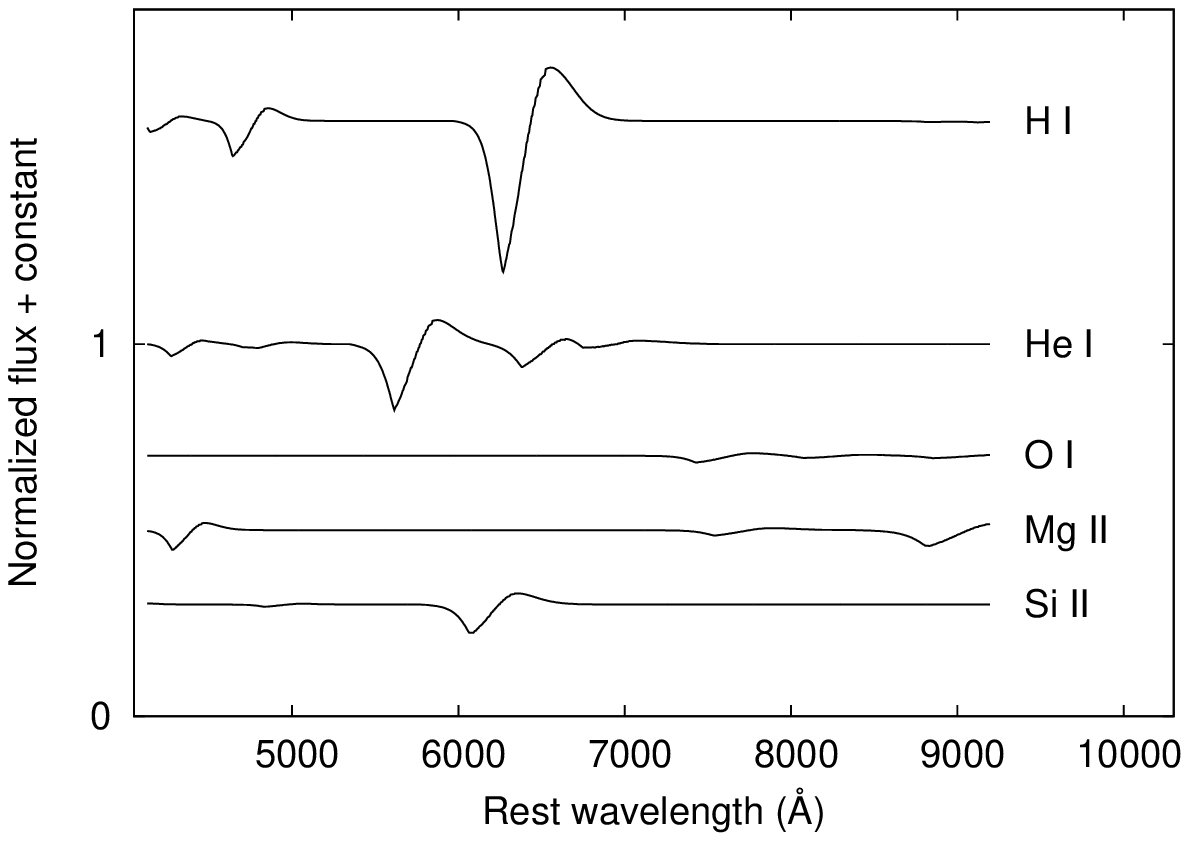} \hspace{5mm}
\includegraphics[width=.45\textwidth]{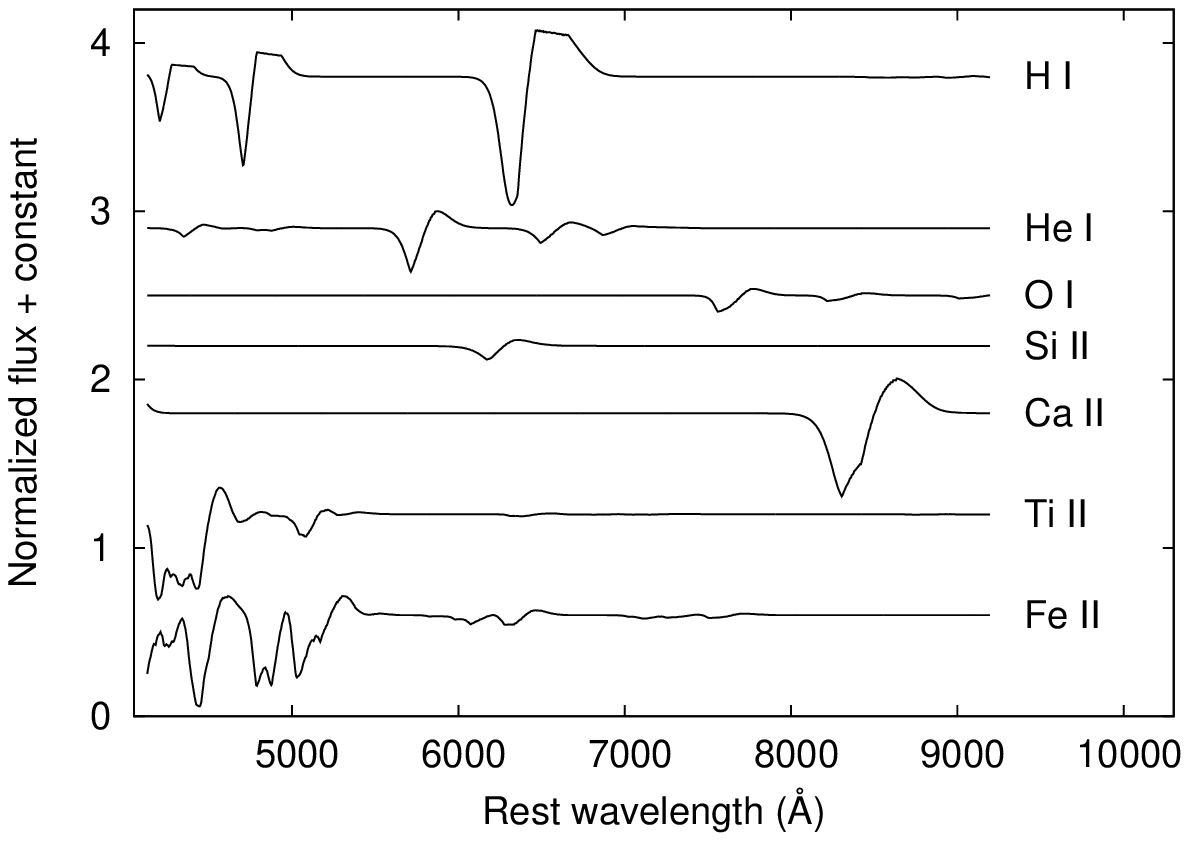}
\caption{Best-fit SYNAPPS models of two selected spectra (+9 and +27 days), and the contribution of the single elements to the model spectra.}
\label{fig:sp_model}
\end{center}
\end{figure*}

Our SYNAPPS modelling shows a continuously decreasing photosperic temperature from 8000 K (+9 d) to 5800 K (+34 d). This agrees well with the results
of both the spectral modelling of BA15 and the blackbody-fitting of the spectral continua carried out by MG14, as well as with the results concerning 
the modelling of SN~1993J \citep{Elmhamdi06}. The photosperic velocity drops in our models from 14\,100 km s$^{-1}$ to 8000 km s$^{-1}$ during the first month
after explosion (see Table \ref{tab:synpar}). These values are significantly larger than those published by BA15 (9600 km s$^{-1}$ at +13d and 5900 km s$^{-1}$ at +26d), but are in closer agreement with 
the results of \citet{Elmhamdi06} for SN~1993J (9000 km s$^{-1}$ at +16d and 8000 km s$^{-1}$ at +24d).

Best-fit SYNAPPS models of two selected spectra (+9 and +27 days), and the contribution of the single 
elements to the model spectra are shown in Fig. \ref{fig:sp_model}. As the bottom panels show, H\,\begin{small}I\end{small}, 
He\,\begin{small}I\end{small}, Si\,\begin{small}II\end{small}, and Mg\,\begin{small}II\end{small} 
are the most abundant ions in the ejecta in the earliest phase, while the Ca\,\begin{small}II\end{small} NIR triplet,
the O\,\begin{small}I\end{small} $\lambda 7774$ line, as well as Fe\,\begin{small}II\end{small} and Ti\,\begin{small}II\end{small} features become strong after +19d (which is around the epoch of the optical 
photometric maximum, see Figure \ref{fig:lc}). Based on our modelling, the $\sim$4200--5000 \AA\ region in the post-maximum spectra may be a complex blend of H$\gamma$, 
He\,\begin{small}I\end{small} $\lambda 4471$, Fe\,\begin{small}II\end{small} and Ti\,\begin{small}II\end{small} lines. This agrees well with the line
identifications of BA15, except that they found Co\,\begin{small}II\end{small} being dominant instead of Ti\,\begin{small}II\end{small} in this region.

Since it is always a key point in the studies of Type IIb/Ib/Ic SNe, we also examined the evolution of H\,\begin{small}I\end{small} and 
He\,\begin{small}I\end{small} lines in detail. As can be seen in Figures \ref{fig:sp} and \ref{fig:sp_model}, there is a broad, two-component 
absorption profile at $\sim$6200 \AA. This is a common feature in the photosperic spectra of SN IIb \citep{Matheson00,Matheson01} and SN Ib 
\citep[see e.g.][and references therein]{Wheeler94,Branch02,Folatelli06,Hachinger12,Reilly16}. While the red component of this absorption feature belongs obviously to the P Cygni profile of the 
H$\alpha$ line, the origin of the blue component cannot be determined unambiguously. Several authors have suggested that the blue component also belongs to the H$\alpha$  
line, e.g. due to the presence of a non-spherical density distribution of H \citep[see e.g.][MG14]{Schmidt93}, or of a second, outer layer of H 
producing high-velocity (HV) H$\alpha$ lines \citep[see e.g.][]{Zhang95,Branch02,Folatelli14,Marion14}.
At the same time, the blending of photospheric H$\alpha$ with other ions, e.g. C\,\begin{small}II\end{small} $\lambda 6580$ or more likely Si\,\begin{small}II\end{small} 
$\lambda 6355$ \citep[see the latter case e.g. in][]{Mazzali09,Silverman09,Hachinger12} can be also an alternative explanation. 

\begin{table}
\begin{center}
\caption{The photospheric velocities and temperatures of SN~2013df, as found by SYNAPPS}
\label{tab:synpar}
\begin{tabular}{cccc}
\hline
\hline
Date & Epoch & $v_\rmn{phot}$ [km s$^{-1}$] & $T_\rmn{phot}$ [K] \\
\hline
2013-06-13 & +9 & 14\,100 & 8000 \\
2013-06-16 & +12 & 11\,800 & 7600 \\
2013-06-23 & +19 & 10\,800 & 6800 \\
2013-06-27 & +23 & 8450 & 6500 \\
2013-07-01 & +27 & 8660 & 6000 \\
2013-07-08 & +34 & 8000 & 5800 \\
\hline
\end{tabular}
\end{center}
\end{table}

In the case of SN~2013df, we suggest that the presence of Si\,\begin{small}II\end{small} is a viable option to explain the source of the blue component of
the broad absorption profile at 6200 \AA. In our SYNAPPS models we were able to fit this feature with Si\,\begin{small}II\end{small} at each epoch
using the temperatures and photosperic velocities mentioned above. We found that the line strength of Si\,\begin{small}II\end{small} $\lambda 6355$ 
decreased continuously in time, similarly to the cases of SN~2001ig \citep{Silverman09} and SN~2003bg \citep{Mazzali09}.
Radiative transfer simulations of SN IIb/Ib/Ic spectra also predict the presence of freshly synthesized Si in the photospheric phase \citep{Dessart11,Hachinger12}; 
however, \citet{Hachinger12} note that in more H-rich cases (i.e. SN IIb) the HV wing of H$\alpha$ may dominate over the Si\,\begin{small}II\end{small} $\lambda 6355$ profile.
In the case of SN~2013df, we did not find any signs of HV H$\beta$ lines, which may be an argument against the presence of HV H in 
the outer layers of the ejecta. At the same time, as \citet{Marion14} described in their study of SN~2011dh, it is possible that the presence of an outer H 
region does not produce HV lines except in the case of H$\alpha$.
As a conclusion, we do not rule out that this puzzling feature corresponds to HV H, but we suggest that its source is more probably Si\,\begin{small}II\end{small}. 

Another factor that makes the identification of the $\sim$6200 \AA\ profile ambiguous is that there are some difficulties in fitting the P Cygni profile of H Balmer lines in 
the spectra of SN~2013df with SYNAPPS. On the one hand, H$\alpha$ and H$\beta$ absorption lines cannot be fitted contemporaneously with a given set of parameters of H lines 
(in Fig. \ref{fig:sp_model}, we present the solution where H$\beta$ line is well-fitted).
On the other hand, there is also a general inconsistency in the fitting of the absorption and emission components of the H$\alpha$ line.
The same problem was described by \citet{Elmhamdi06} and \citet{Kumar13} concerning the SYNOW modelling of the spectra of
SNe~1993J and 2011fu, respectively, and can be also seen in the results of \citet{Sahu13} concerning SYN++ modelling of some pre-maximum spectra of SN~2011dh.
This modelling problem may be caused by a still unrevealed blending effect, or, more probably, by the fact that one of the basic assumptions of SYNOW and SYNAPPS codes is local thermodynamic equilibrium (LTE)
for level populations. While the presence of non-LTE effects seems to be a plausible explanation for the differences of the observed and synthetic spectra, we note that BA15 were also unable to fit the 
P Cygni profile of H$\alpha$ adequately, although they used a radiative-transfer code that includes abundance stratification and a module that calculates the ionization of H
and He in full non-LTE (see references in BA15).

To get the best-fit models of the post-maximum spectra, the H lines need to be detached from the photosphere by $\sim$1000 km s$^{-1}$.
This results flat-topped H emission components in the model of the +27d spectrum; 
however, we were not able to check whether this effect can be real or not. The emission components of H$\alpha$ and H$\beta$ are blended with the He\,\begin{small}I\end{small} $\lambda 6678$ line and metal 
lines, respectively, and the inconsistencies concerning the contemporaneous fit of H$\alpha$ and H$\beta$ lines makes this examination even more difficult.
Nevertheless, we note that 1000 km s$^{-1}$ is at the limit of uncertainty in SYNAPPS fits, so this may not be a real separation.

The other interesting topic is the evolution of He lines. As can be seen in Figs \ref{fig:sp} and \ref{fig:sp_model}, as well as in other SN IIb, 
He\,\begin{small}I\end{small} $\lambda 5876$ is already identifiable in the pre-maximum spectra, and its intensity is continuously growing. While this feature 
is often referred as to a blend of He\,\begin{small}I\end{small} and Na\,\begin{small}I\end{small} D, our modelling shows that the contribution of Na\,\begin{small}I\end{small} 
to this feature is negligible (see the details later).

We also identified other He\,\begin{small}I\end{small} lines in the spectra of SN~2013df. In our HET spectra, the $\lambda$6678 and $\lambda$7065 lines are identifiable as early as 
+9d and +19d, respectively, in contrast to the results of MG14 who stated that these lines are not visible before +40d (using June 4.3 UT as the explosion date). The $\lambda$5016 line 
becomes visible at +19d. The He\,\begin{small}I\end{small} $\lambda$4471 line, although it is the strongest He line after $\lambda$5876, is not obviously identifiable in our HET 
spectra. On the one hand, the blue end of the observed spectral region is at 4170 \AA, very close to the blue wing of the suggested He\,\begin{small}I\end{small} line. On the other hand, 
the observed feature at $\sim$4200$-$4300 \AA\ may be a complex blend of He\,\begin{small}I\end{small}, H$\gamma$ and Mg\,\begin{small}II\end{small}, or, of He\,\begin{small}I\end{small}, 
H$\gamma$, Ti\,\begin{small}II\end{small} and Fe\,\begin{small}II\end{small} (before and after maximum, respectively). 

We also note that we did not find convincing evidence for non-thermal excitation of the He
features during the early phases of SN 2013df sampled by our spectra, contrary to
the findings of \citet{Ergon14} in SN~2011dh. Such an effect is usually invoked
to explain the increase of the optical depth of the HeI features as a function of
phase, contrary to the expectations from the Sobolev approximation.

\subsection{``High-velocity'' helium in the ejecta}\label{sp_he}

\begin{figure}
\begin{center}
\includegraphics[width=.5\textwidth]{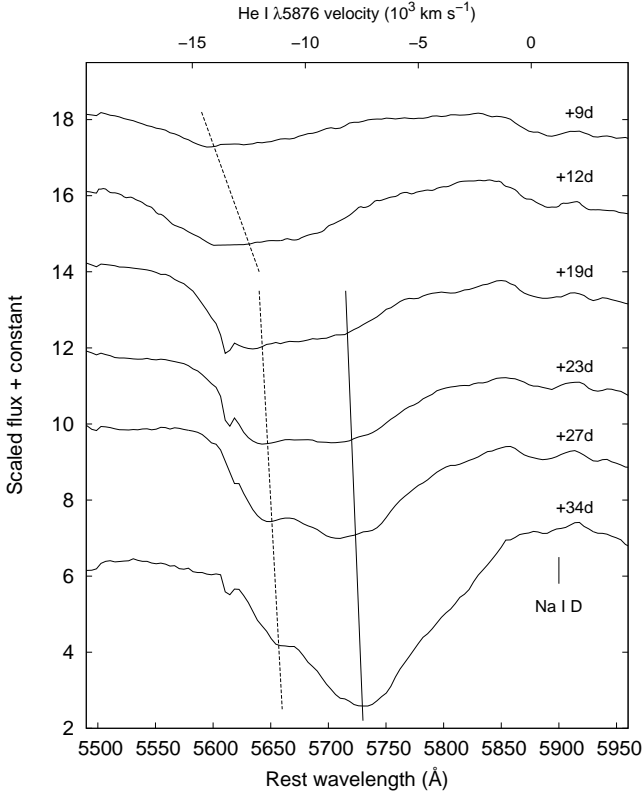}
\caption{Evolution of the He I $\lambda$5876 line in the spectra of SN~2013df. Solid and
dashed vertical lines mark the positions of ``low''- and ``high-velocity'' He I features, respectively.
The weak feature at $\sim$5900 \AA\ is the unresolved sum of interstellar Na I D lines originating from the Milky Way and the host.}
\label{fig:5876}
\end{center}
\end{figure}

During the analysis of the evolution of He\,\begin{small}I\end{small} lines, we discerned that the absorption profile at $\sim$5650 \AA, which we first identified as the 
He\,\begin{small}I\end{small} $\lambda$5876 line, actually consists of two components. As Fig. \ref{fig:5876} shows, there seems to be a 
single absorption line with a continuously decreasing velocity in the pre-maximum phases (up to +19d); however, as the spectra evolves, it develops a double-troughed
feature. The red component becomes dominant by +34 day. 

While the origin and evolution of the double-troughed 6200 \AA\ profile are discussed in several papers on SN IIb, the similar structure of the 
5650 \AA\ profile does not seem to be a common phenomenon.
Up to now, the only discussion of the double-troughed structure of the He\,\begin{small}I\end{small} $\lambda$5876 line was for SN~1993J \citep{Schmidt93, Zhang95}. 
Based on the explanation of \citet{Schmidt93}, the shape of this feature, as well as the similar double-troughed profile of the H$\alpha$ line, may be a sign of asymmetry in the expanding material.
The presence of this asymmetry is also supported by the results of spectropolarimetric studies of SN~1993J \citep{Trammell93,Hoeflich95,Hoeflich96,Tran97}; similar findings have been published later 
based on spectropolarimetry of several other SN IIb \citep[see][and references therein]{Mauerhan15}, as well as the on the high-resolution imaging of Cas A and the spectral studies of its 
light echoes \citep[see e.g.][]{DeLaney10,Rest11}.

\citet{Zhang95} also studied the spectroscopic evolution of the broad H$\alpha$ and He\,\begin{small}I\end{small} $\lambda$5876 absorption features. Instead of assuming an asymmetric explosion, they 
suggested a model with a two-component density structure of the ejecta (a shallower layer immediately outside a steeply decreasing inner envelope) to explain the shape of H and He features. 
Using this two-component model, they were able to reproduce the H and He line profiles better than in the cases when they used models based on a single power-law density stucture.

\begin{figure*}
\begin{center}
\includegraphics[width=.45\textwidth]{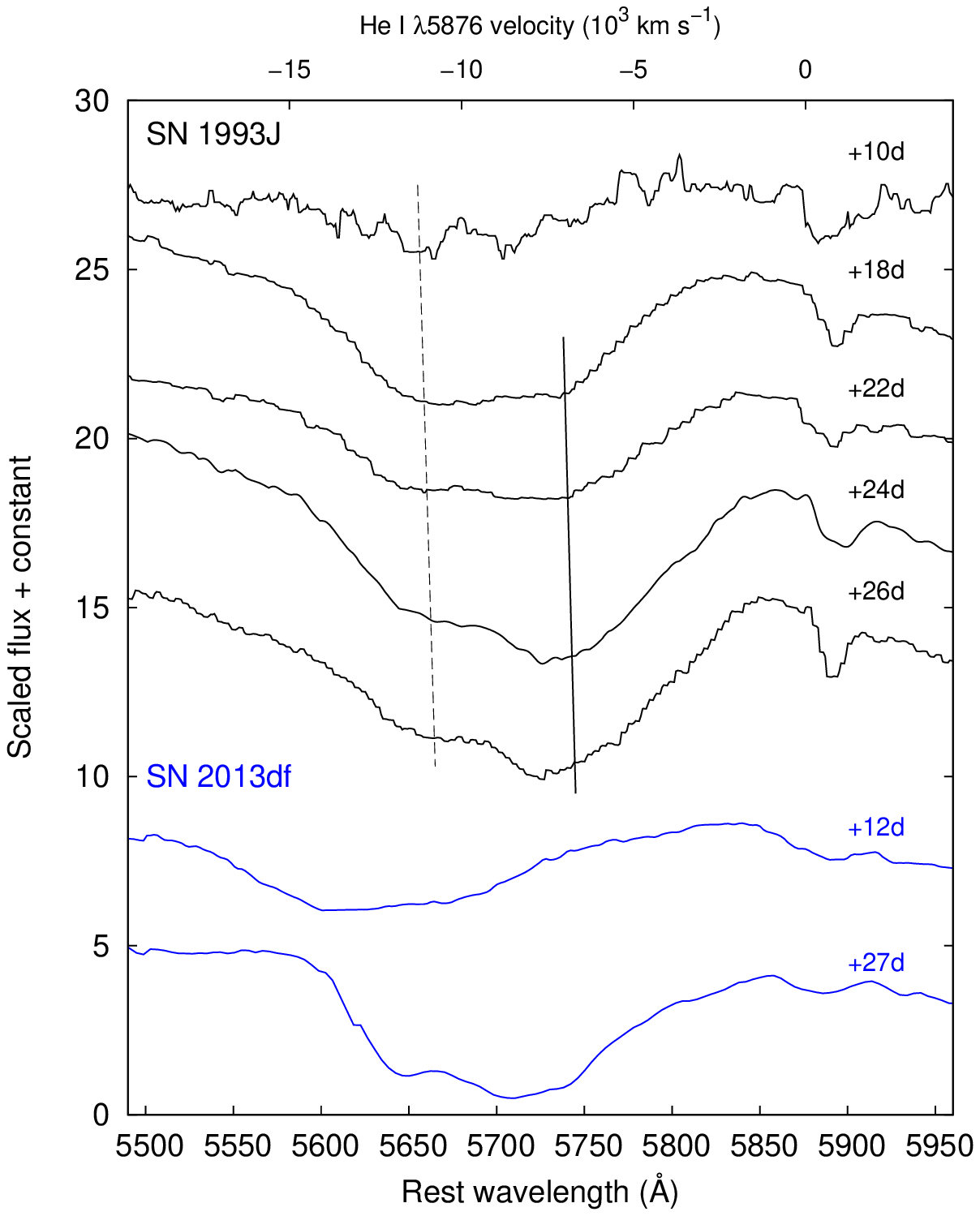} \hspace{5mm} \vspace{5mm}
\includegraphics[width=.45\textwidth]{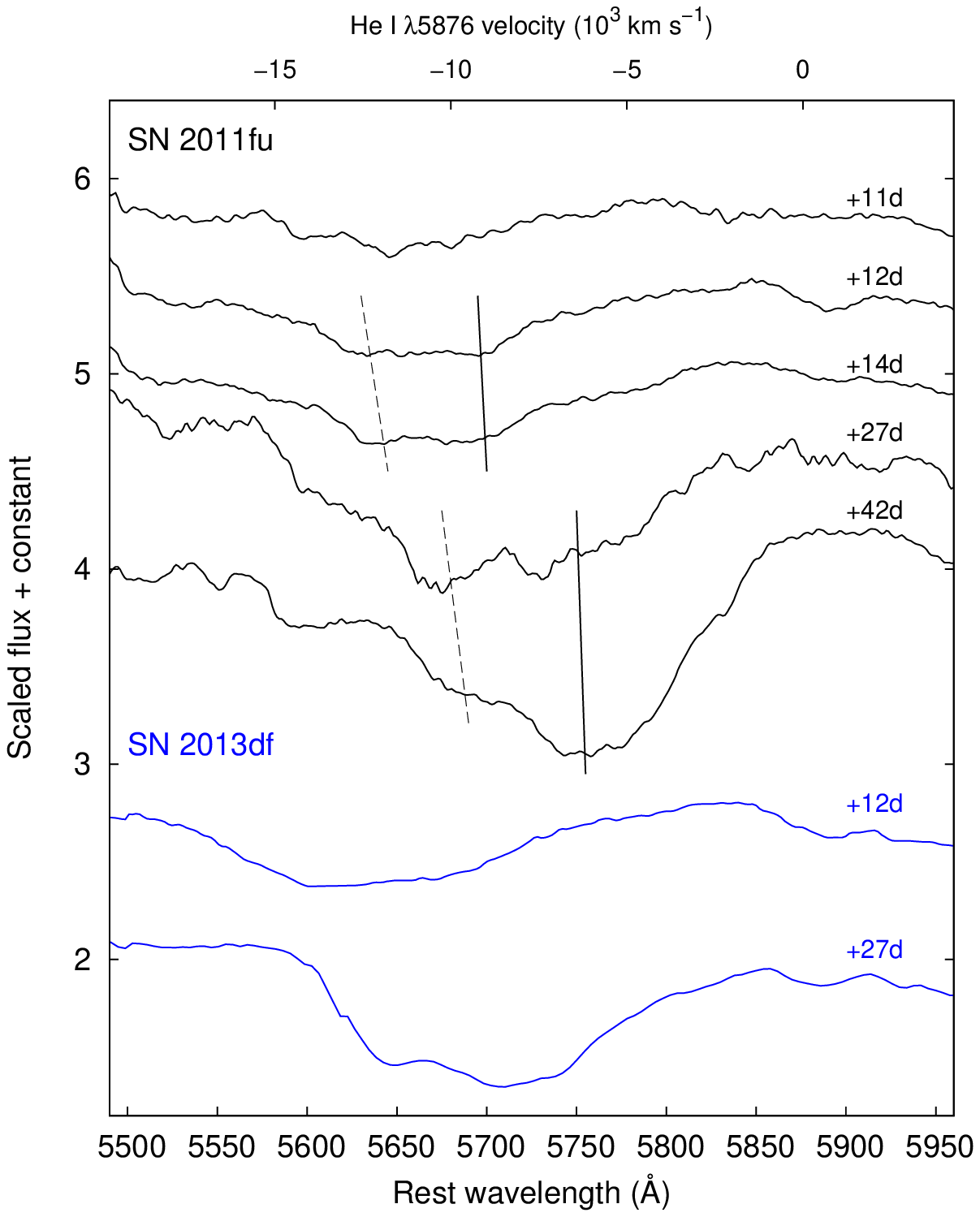}
\includegraphics[width=.45\textwidth]{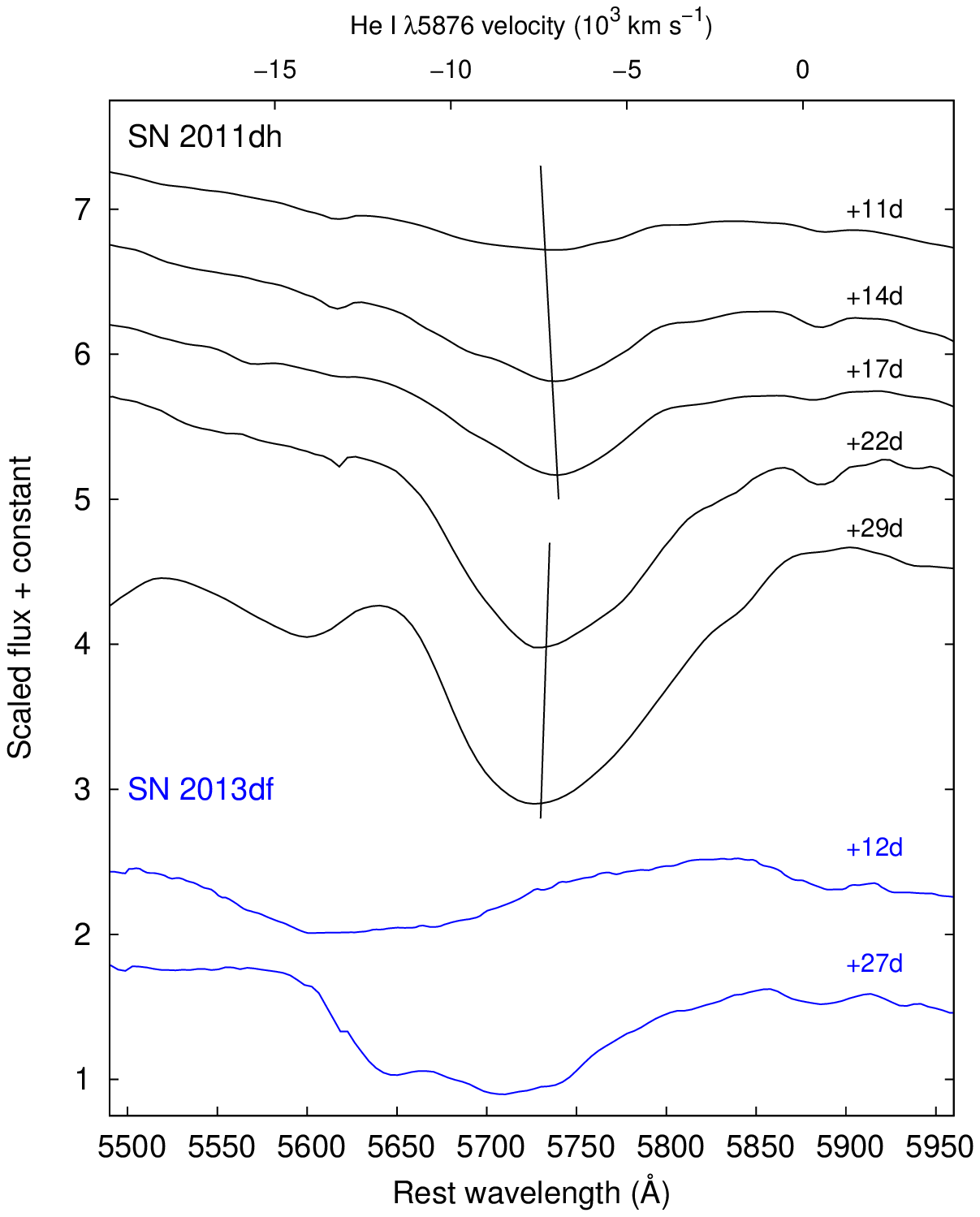} \hspace{5mm}
\includegraphics[width=.45\textwidth]{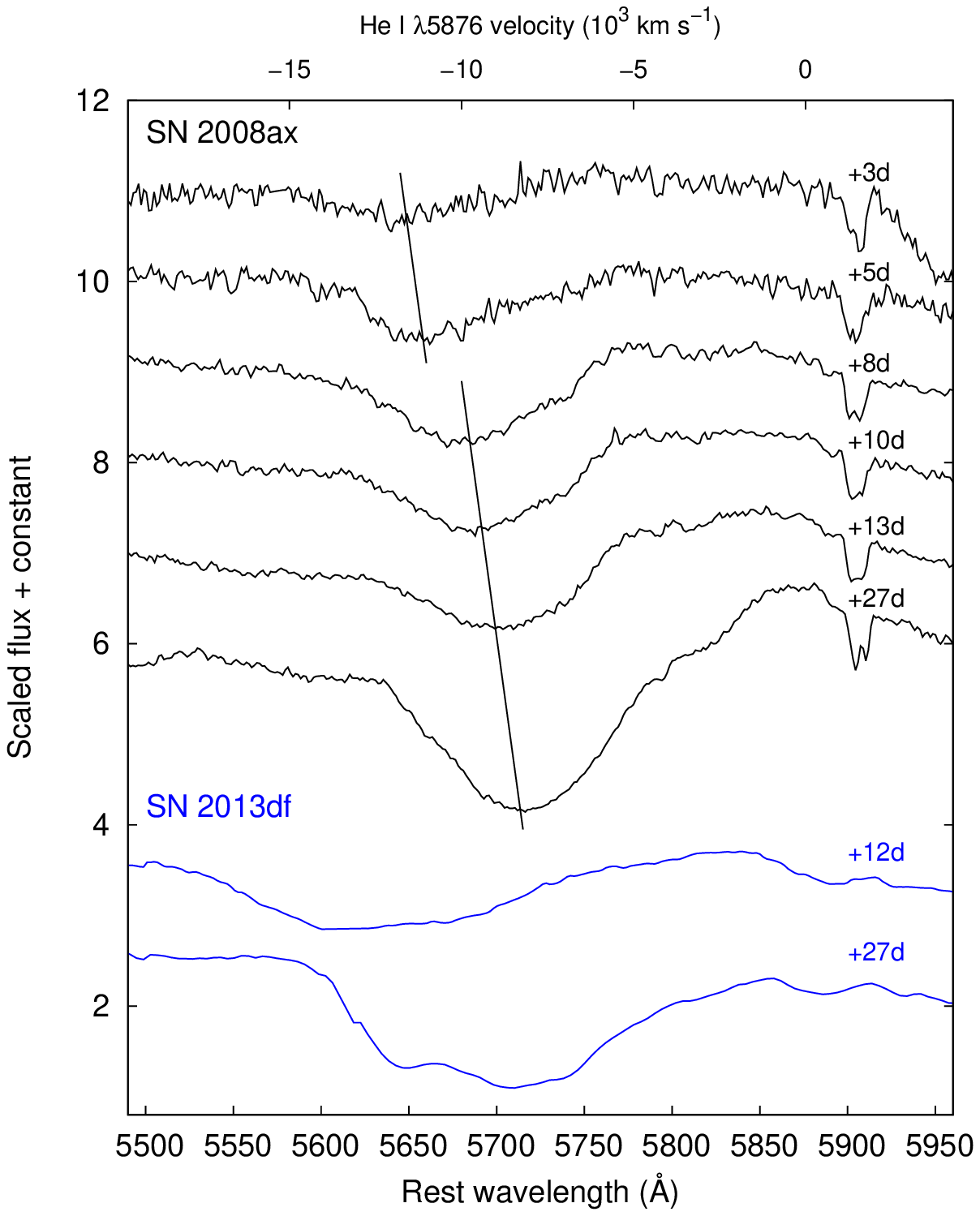}
\caption{Spectral evolution of four different SN~IIb between 5500--5950 \AA: SN~1993J \citep[][{\it top left}]{Barbon95}, SN~2011fu \citep[][{\it top right}]{Kumar13}, 
SN~2011dh \citep[][{\it bottom left}]{Marion14}, and SN~2008ax \citep[][{\it bottom right}]{Taubenberger11,Modjaz14}. The spectra of SNe~1993J and 2011fu show a double-troughed 
He I $\lambda$5876 profile similarly to SN~2013df, while there are no obvious similar effects in the other two cases shown in the bottom panels. 
Solid and dashed vertical lines mark the positions of ``low''- and ``high-velocity'' He I features, respectively. The +12d and +27d spectra of SN~2013df are also shown on each panel for a comparison.
All spectra were corrected to the redshifts of the host galaxies. The spectra of SN~2011fu were smoothed by a 20\AA-wide window function for better visibility. All the redshift and age information was 
adopted from the original papers.}
\label{fig:93J_he}
\end{center}
\end{figure*}

The top left panel of Figure \ref{fig:93J_he} shows the series of spectra of SN~1993J \citep[published by][]{Barbon95} in the region of 5500--5950 \AA\ from +10 to +26 days after 
explosion. The double-troughed structure of the 5650 \AA\ profile is clearly seen, and is very similar to that of SN~2013df.

Another SN IIb that seems to show a double-troughed 5650 \AA\ profile is SN~2011fu. \citet{Kumar13} did not study this profile in detail; however,
it is not so easy to see the phenomenon in their spectra because of the low signal-to-noise ratio of the data and the lack of observations between +14 and +27 days after explosion.
We present here the earliest five spectra from their paper in the top right panel of Figure \ref{fig:93J_he}; the 5650 \AA\ profile clearly has a double-troughed shape at +27d
(the data were smoothed here by a 20\AA-wide window function for better visibility). 
\citet{MG15} presented high-quality spectra of SN~2011fu including data taken at +17, +20 and +27 days. While they noted that around a
month past explosion the He\,\begin{small}I\end{small} $\lambda$5876 absorption component had a complex profile with a double trough, they did not analyze the profile in detail.

We also checked all the available spectra of other SN IIb. We did not see this double minimum either in our data of SN~2011dh \citep[published in][]{Marion14} or
in any other cases where the first 30-40 days are well sampled\footnote{We downloaded these data from WISeREP (Weizmann Interactive Supernova data REPository), 
http://www.weizmann.ac.il/astrophysics/wiserep, \citet{Yaron12}.}: SNe 1996cb, 1998fa \citep[both from][]{Modjaz14},
2000H \citep{Branch02,Modjaz14}, 2001ig \citep{Silverman09}, 2003bg \citep{Hamuy09}, 
2005U, 2006T, 2006el, 2008aq \citep[all from][]{Modjaz14}, 2008ax \citep{Taubenberger11,Modjaz14}, 2008bo, 2008cw \citep[both from][]{Modjaz14}, 2011hs \citep{Bufano14}, and 2013bb 
(unpublished PESSTO\footnote{Public ESO Spectroscopic Survey of Transient Objects, www.pessto.org} data).
We found three other SN IIb with well-sampled spectroscopic datasets: SNe~2009mg \citep{Oates12}, 2010as \citep{Folatelli14}, and 2011ei \citep{Milisavljevic13}.
These data are not publicly available; however, the published data seem to show the double-troughed 5650 \AA\ profile in none of these cases.
We present the early spectral evolution of two of the listed SNe, 2008ax and 2011dh, in the bottom panels of Figure \ref{fig:93J_he}.

\begin{figure}
\begin{center}
\leavevmode
\includegraphics[width=.5\textwidth]{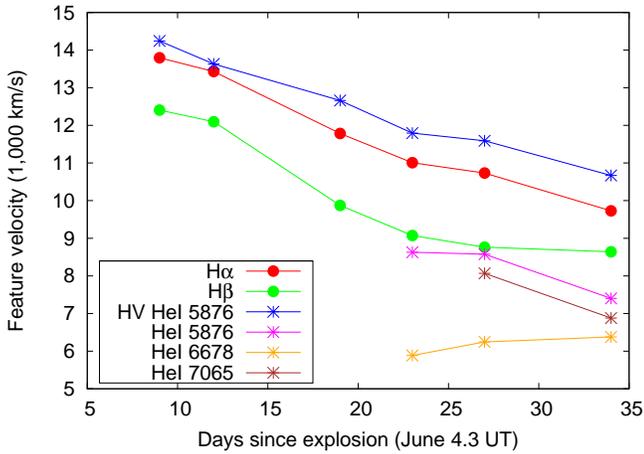}
\caption{Velocities of some selected lines in the photosperic phase of SN~2013df. Note that the velocity of the He I $\lambda$6678 line has a large uncertainty at 
each epoch; that may explain why its velocity evolution is so different from those of other He lines.}
\label{fig:vel}
\end{center}
\end{figure}

While \citet{Schmidt93} and \citet{Zhang95} suggested that the double-troughed appearance of the $\sim$5650 \AA\ feature may be caused by the asymmetry of the explosion or the spatial distribution 
of He in the ejecta, respectively, we also examined other possibilities.
This feature is generally interpreted as the unresolved blend of 
He\,\begin{small}I\end{small} $\lambda$5876 and Na\,\begin{small}I\end{small} D lines. Nevertheless, we think that Na\,\begin{small}I\end{small} D probably does not interfere here. 
Looking at Figure \ref{fig:5876}, we can see that the velocity of the $\sim$5650 \AA\ feature is $\sim$14\,000 km s$^{-1}$ with respect to the rest wavelength of 
He\,\begin{small}I\end{small} $\lambda$5876 at the first epoch. Na\,\begin{small}I\end{small} D has higher rest wavelength, so if the feature at $\sim$5650 \AA\ is Na\,\begin{small}I\end{small} 
instead of He\,\begin{small}I\end{small}, it should have even higher velocity than 14,000 km s$^{-1}$ (which value matches pretty well with the velocity of H$\alpha$ 
and with the global photosperic velocity determined by SYNAPPS, see Figure \ref{fig:vel} and Table \ref{tab:synpar}, respectively).
Later, when the lower-velocity components take over, the assumption that it is Na\,\begin{small}I\end{small} instead of He\,\begin{small}I\end{small} would, 
again, imply that Na\,\begin{small}I\end{small} is at $\sim$8,500-9000 km s$^{-1}$, which is still a higher value than those of the rest of the photospheric features. 

\begin{figure}
\begin{center}
\includegraphics[width=.5\textwidth]{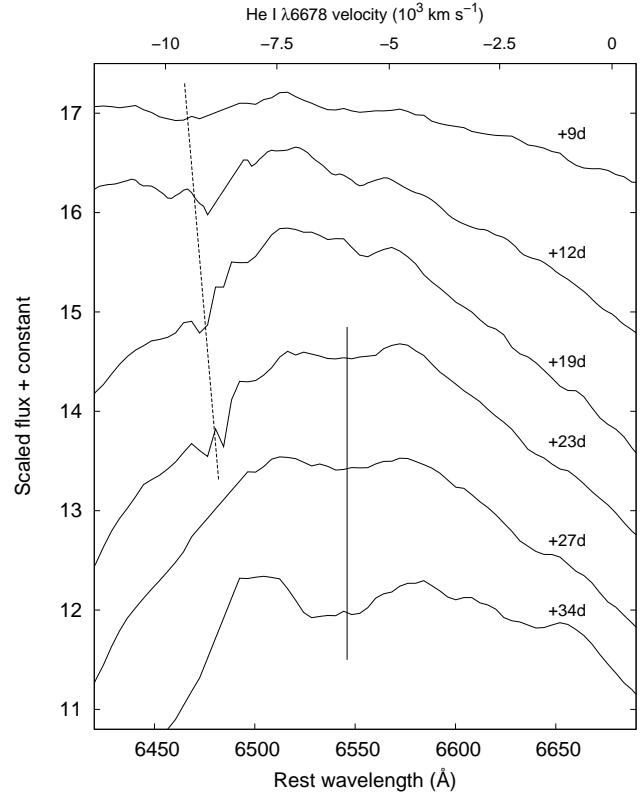}
\caption{Spectral evolution of the He I $\lambda$6678 line in the case of SN~2013df. Solid and
dashed vertical lines mark the positions of ``low''- and ``high''-velocity He I features, respectively.}
\label{fig:6678}
\end{center}
\end{figure}

Our SYNAPPS models also suggest that the contribution of local Na\,\begin{small}I\end{small} to the spectra of SN~2013df is negligible (we note that this is also in 
agreement with the modelling results of BA15). Since our models adequately fit
the spectra, and we did not find any other elements that have lines around 5650 \AA\ (see Figure \ref{fig:sp_model}), we assume, as do \citet{Schmidt93} and \citet{Zhang95}, that 
both absorption features belong to the He\,\begin{small}I\end{small} $\lambda$5876 line.
To verify this statement, we also examined the evolution of other He lines in the spectra of SN~2013df. As we mentioned above, the other He lines in the observed spectral range are all 
weak and/or blended; however, if we take a closer look at He\,\begin{small}I\end{small} $\lambda$6678 at the top of the emission of the H$\alpha$ line, we can see a weak double-troughed profile 
evolving similarly as $\lambda$5876 (see Fig. \ref{fig:6678}).

\begin{figure}
\begin{center}
\leavevmode
\includegraphics[width=.5\textwidth]{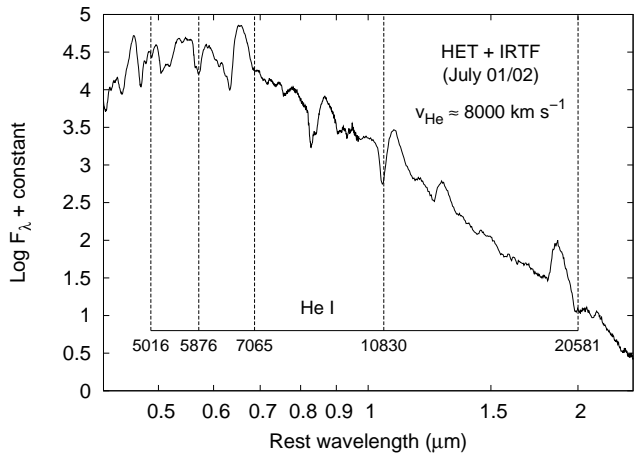}
\caption{The combined optical/NIR spectrum of SN 2013df at +27/28 d after explosion. The positions of the most significant He I lines (all at $v \sim$8000 km s$^{-1}$) are marked with 
dashed lines.}
\label{fig:NIR}
\end{center}
\end{figure}

An independent check on the behavior of the He lines would be to examine He\,\begin{small}I\end{small} lines in the NIR spectral range ($\lambda\lambda$ 10830, 20581). Unfortunately, we have 
only a post-maximum NIR spectrum, thus, we were not able to follow the evolution of these He\,\begin{small}I\end{small} lines during the photospheric phase, as well as to check whether they 
also show double-troughed profiles in the early phases or not. Our single NIR spectrum was obtained nearly contemporaneously with the +27d optical spectrum (we show the combined optical/NIR spectrum 
in Figure \ref{fig:NIR}). As can be seen, the NIR He\,\begin{small}I\end{small} lines are found at $v \sim$8000 km s$^{-1}$, which agrees well with the velocities of the optical 
He\,\begin{small}I\end{small} lines at this epoch (see Figure \ref{fig:vel}). Note that the velocity of the He\,\begin{small}I\end{small} $\lambda$6678 line has a large uncertainty at each epoch; 
that may explain why its velocity evolution is so different from those of other He lines.

As discussed e.g. in \citet{Marion14}, the hydrogen velocities usually significantly exceed the helium velocities in the case of either SN Ib \citep{Branch02} or SN IIb 
\citep[see also][]{Chornock11,Milisavljevic13}; the degree of separation can be as large as 4-5000 km s$^{-1}$. The presence of this separation also agrees with the results of the 
theoretical work by \citet{Dessart11}. The clear separation of H and He lines in velocity space indicates that the lines of the two elements are formed in separate regions.

Assuming that we see the evolution of He\,\begin{small}I\end{small} at $\sim$5650 \AA, Figure \ref{fig:vel} allows us to draw an interesting conclusion.
As can be seen in the case of SN~2013df, the ``high-velocity'' He\,\begin{small}I\end{small} $\lambda$5876 component has a similar (actually even a bit higher) 
velocity to that of the H$\alpha$ line.
Instead of the model described above, this effect indicates that both H and He features form in the outermost envelope during the early phases.
The evolution of the He\,\begin{small}I\end{small} $\lambda$5876 line is consistent with the nature of a two-component atmosphere, which consists of an outer H-rich shell mixed with some He 
and a denser He-rich core. The transition in the line profiles occurs when the photosphere moves back from the shell to the innner core.
The mixing of H and He in the outer shell is in agreement with the modelling reults of BA15. MG14 found the H and He velocities to be separated by $\sim$1000-1500 km s$^{-1}$; 
however, they do not mention the presence of ``high-velocity'' He in the spectra.
We note that the velocity of the the ``high-velocity'' He\,\begin{small}I\end{small} $\lambda$5876 component is among the highest He line velocities that have ever been observed in SNe IIb. 
It is comparable to those of SNe Ib \citep[see e.g. Figure 3 of][]{Liu15}, which tend to show higher He line velocities.

We expected to find similar results concerning both SNe~1993J and 2011fu. In the case of SN~1993J, the results of \citet{Barbon95} show a quite large separation ($\sim$4000 km s$^{-1}$) between the H$\alpha$ and 
He\,\begin{small}I\end{small} $\lambda$5876 lines. At the same time, if we take into account the presence of the ``high-velocity'' He component ($v \sim$ 11-12\,000 km s$^{-1}$), the degree 
of separation is less than 2000 km s$^{-1}$ \citep[as was previously indicated by the results of][and, more recently, by BA15]{Wheeler93}.
In the case of SN~2011fu, \citet{MG15} examined the velocity evolution of H$\alpha$ and He\,\begin{small}I\end{small} $\lambda$5876 lines in detail. Although they found the He\,\begin{small}I\end{small} $\lambda$5876 line 
to have a complex profile, they obtained the line velocities by adjusting a single Gaussian to the whole profile. Nevertheless, they found that H$\alpha$ is not clearly separated from He in velocity space, which is in agreement 
with our findings concerning SNe~2013df and 1993J.

The results of \citet{Rest11}, based on the analysis of light echo spectra of Cas A, are also worth mentioning here. They found the H and He velocities being coincident, for which they gave several possible explanations: 
a relatively strong mixing of H and He layers, the role of the distribution of $^{56}$Ni in the ionization structure of the outer layers, or the extreme thinness of the outer H layer. Although there is no clear evidence as to 
whether this SN was a cIIb or an eIIb, the second option (assuming a red supergiant as progenitor) seems to be more probable. The explosion of a compact star can be a viable explanation only in binary progenitor models, 
but there is no evidence for a companion star to date; however, the merging of two stars into a single one before the explosion may solve this problem \citep{Young06,Krause08,Claeys11}.
If we suppose the scenario of a single extended progenitor for Cas A, the findings of \citet{Rest11} fits well into the results described above. While they did not rank the possible explanations of the coinciding H and He velocities, 
we think that the extreme thinness of the outer H layer is less possible; in this case, we should see this effect also in SNe cIIb.

As a conclusion, it can be pointed out that the evolution of the He\,\begin{small}I\end{small} $\lambda$5876 profile in the spectra of SN~2013df, as well as the lack of considerable separation of H and He velocities indicates 
that He lines partially form in the outer regions of the expanding ejecta. Similar spectral properties can be seen in SNe~1993J and 2011fu, as well as in the light echo spectra of Cas A. All of these SNe probably belong to SN eIIb 
that are thought to have very extended progenitors. On the other hand, as we found, many other SN IIb, which are thought to emerge from more compact progenitors, do not show the presence of ``high-velocity'' He lines.
The presence of this effect may depend on the degree of the mixing of H and He layers, which may be more significant in red/yellow supergiants than in more compact Wolf-Rayet stars; however, there are several circumstances that may 
affect the degree of mixing \citep[convection, rotation, presence of a companion etc., see for a review][]{Langer12}. All of these factors should be taken into account in a detailed examination of the problem, which is beyond the 
scope of our paper. For the more detailed studies, it would be also necessary to obtain and analyse high-quality and well-sampled spectral data of other SNe eIIb and cIIb.

\section{Light curve analysis}\label{lc}

\subsection{Properties of the early-time light curves}\label{lc_lc}

\begin{figure}
\begin{center}
\includegraphics[width=.45\textwidth]{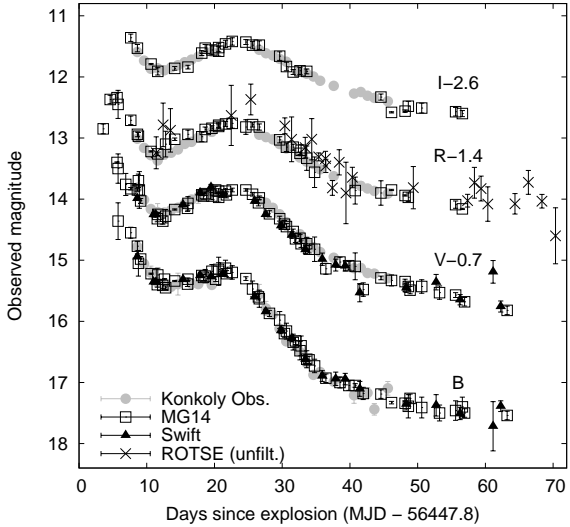}
\caption{Our {\it BVRI} measurements of SN~2013df comparing with the results of \citet[][MG14]{MG14} and with Swift data. Our early-time unfiltered ROTSE measurements are also shown.}
\label{fig:lc_early_comp}
\end{center}
\end{figure}

MG14 carried out a detailed UV-optical photometric study of SN~2013df, including the analysis of light-curve shapes, colour curves, and the bolometric light curve.
Since our data agree well with the observed brightness of the SN published by MG14 (see Figure \ref{fig:lc_early_comp} showing our {\it BVRI} data comparing with the results 
of MG14 and with Swift data), we have not repeated every step of the analysis of MG14. Instead, in this section we present some additional results that complement 
the formerly published ones, or, in some cases, may lead to different conclusions.

SN~2013df is only the third SN IIb, after SNe~1993J \citep{Richmond94} and 2011fu \citep{Kumar13,MG15}, where the initial declining phase is clearly visible in all optical bandpasses.
This effect may be related to the cooling of the extended progenitor envelope after shock-breakout. Similar light curve evolution seems to be detected in the Sloan {\it g'} bandpass in the case 
of SN~2011dh within some days after explosion \citep{Arcavi11,Bersten12}. Other SN IIb caught in very early phases do not show this effect \citep[see e.g. SN 2013cu,][]{GalYam14}, 
or only in the UV and blue bandpasses \citep[see the case of SN~2008ax,][]{Pastorello08,Roming09}. As discussed in Section \ref{intro}, the detectability of the initial declining phase 
is thought to be connected with the radius of the progenitor star.
 
SN IIb also show significant inhomogeneity in their secondary peak magnitudes and in the corresponding peak times. We present these parameters for SN~2013df in Table \ref{tab:lcmax}, and 
compare them with those of three other SNe IIb (1993J, 2008ax, and 2011dh) in Table \ref{tab:lcmax_comp}. For SNe 2008ax and 2011dh, we adopted the results of \citet{Taubenberger11} and
\citet{Sahu13}, respectively. We re-calculated the peak absolute magnitudes of SN~1993J published by \citet{Richmond94}, because the value of the interstellar reddening 
in the direction of SN~1993J was very uncertain at the date of publishing (our adopted values of reddening, the distance moduli, and the estimated dates of explosions, together with the 
corresponding references, are presented in the footnotes of Table \ref{tab:lcmax_comp}).

Our results show that SN~2013df was less luminuous than SN~1993J. On the other hand, SN~2013df seems to be the spectroscopic ``twin'' of SN~1993J, as mentioned 
above. SN~2013df also seems to be fainter than both SNe~2008ax and 2011dh. There are other faint SN IIb in the literature e.g. SN~1996cb, $M_\rmn{V} \sim$ -16.2 mag \citep{Qui99}, 
or SN~2011ei, $M_\rmn{V} \sim$ -16.0 mag \citep{Milisavljevic13}. The brightest known SN IIb, SN~2011fu, has a peak magnitude of 
$M_\rmn{V} \sim$ -18.5 mag \citep{Kumar13}. We thus see that peak magnitudes scatter within a range of more than 2 magnitudes.
Looking at the values of secondary peak times (the time elapsed from the moment of explosion), the secondary maximum of SN 2013df 
occurred at a later epoch in all bands, relative to that of the other three SNe. These findings agree well with the results published by VD14. 

We note that the results of MG14 suggest a different conclusion. Adopting a significantly larger value for the distance of the host galaxy 
(see Section \ref{obs}), they concluded that SN~2013df was only slightly fainter than SN~1993J ($M_\rmn{R}$ = -17.71 $\pm$ 0.31 mag vs. -17.88 $\pm$ 0.38 mag), so it was brighter than 
both SNe~2008ax and 2011dh. Contrary to MG14, we accept $\mu_0$= 31.10 $\pm$ 0.05 mag for the distance modulus of the host of SN~2013df, as determined by \citet{Freedman01}. This value is close 
to the mean value of the distance moduli in NED ($\mu_0$= 31.23 mag), and it originates from the same study as the distance modulus of M81 (host of SN~1993J) 
adopted by us as well as by MG14 (see Table \ref{tab:lcmax_comp}).

There is also a noticeable difference between the secondary peak times of SN~2013df and the ones determined by MG14. This is caused by the difference between the estimated explosion 
dates (MG14 used $t_0$ = 2,456,450.0 $\pm$ 0.9 JD instead of 2,456,447.8 $\pm$ 0.5 JD). If we correct their results using our $t_0$ value, we get all peak times within 0.5 days in 
the optical ($BVRI$) bands.

\begin{table*}
\begin{center}
\caption{{\it BVRI} and {\it g'r'i'z'} secondary maxima with the corresponding times}
\label{tab:lcmax}
\begin{tabular}{ccccc}
\hline
\hline
Filter & t$_\rmn{rise}^a$ & Apparent max. magnitude & A$_{\lambda}$ & Absolute max. magnitude \\
 ~ & (days) & (mag) & (mag) & (mag) \\
\hline
B & 20.39(0.80)	& 15.31(0.06) & 0.42(0.05) & -16.21(0.12) \\
g' & 20.25(0.22) & 14.86(0.01) & 0.37(0.05) & -16.61(0.08) \\
V & 21.68(0.20)	& 14.55(0.01) & 0.30(0.05) & -16.85(0.08) \\
R & 22.77(0.20)	& 14.26(0.01) & 0.26(0.05) & -17.10(0.08) \\
{\it r'} & 22.38(0.12) & 14.36(0.01) & 0.27(0.05) & -17.01(0.08) \\
i' & 22.78(0.27) & 14.44(0.01) & 0.20(0.05) & -16.86(0.08) \\
I & 23.29(0.20)	& 14.12(0.01) & 0.19(0.05) & -17.17(0.08) \\
z' & 23.02(0.96) & 14.56(0.03) & 0.14(0.05) & -16.68(0.09) \\
\hline
\hline
\end{tabular}
\end{center}
\smallskip
{\bf Notes.} $^a$With respect to $t_{0}$ = 2,456,447.8 $\pm$ 0.5 JD.
\end{table*} 

\begin{table*}
\caption{Comparison of U{\it BVRI} secondary maximum absolute magnitudes and peak times of four SN~IIb}
\label{tab:lcmax_comp}
\begin{tabular}{ccccccccc}
\hline
\hline
~ & \multicolumn{2}{c}{SN 1993J$^a$} & \multicolumn{2}{c}{SN 2008ax$^b$} & \multicolumn{2}{c}{SN 2011dh$^c$ } & \multicolumn{2}{c}{SN 2013df$^d$} \\
Filter & Peak abs. mag. & t$_\rmn{rise}$ & Peak abs. mag. & t$_\rmn{rise}$ & Peak abs. mag. & t$_\rmn{rise}$ & Peak abs. mag. & t$_\rmn{rise}$ \\
~ & ~ & (days) & ~ & (days) & ~ & (days) & ~ & (days) \\
\hline
U & -17.40 $\pm$ 0.14 & 19.0 & -17.73 $\pm$ 0.56 & 16.7 $\pm$ 0.5 & -16.19 $\pm$ 0.18 & 15.8 $\pm$ 0.5 & -- & -- \\
B & -17.23 $\pm$ 0.13 & 19.7 & -17.32 $\pm$ 0.50 & 18.3 $\pm$ 0.5 & -16.38 $\pm$ 0.18 & 19.6 $\pm$ 0.5 & -16.21 $\pm$ 0.12 & 20.39 $\pm$ 0.80 \\
V & -17.59 $\pm$ 0.13 & 21.0 & -17.61 $\pm$ 0.43 & 20.1 $\pm$ 0.4 & -17.12 $\pm$ 0.18 & 20.6 $\pm$ 0.5 & -16.85 $\pm$ 0.08 & 21.68 $\pm$ 0.20 \\
R & -17.75 $\pm$ 0.12 & 21.3 & -17.69 $\pm$ 0.39 & 21.5 $\pm$ 0.4 & -17.43 $\pm$ 0.18 & 21.3 $\pm$ 0.5 & -17.10 $\pm$ 0.08 & 22.77 $\pm$ 0.20 \\
I & -17.68 $\pm$ 0.12 & 22.0 & -17.75 $\pm$ 0.35 & 22.4 $\pm$ 0.7 & -17.48 $\pm$ 0.18 & 22.9 $\pm$ 0.5 & -17.17 $\pm$ 0.08 & 23.29 $\pm$ 0.20 \\
\hline
\hline
\end{tabular}
\smallskip
{\bf Notes.} $^a$Calculated from the data of \citet{Richmond94}; adopted parameters: $t_0$ = 2,449,074.0, $E(B-V)$ = 0.19 $\pm$ 0.09 mag, $\mu_0$ = 27.80 $\pm$ 0.08 mag \citep{Lewis94,Freedman01,Richardson06}.
$^b$Adopted from \citet{Taubenberger11} using $t_0$ = 2,454,528.8 $\pm$ 0.2, $E(B-V)$ = 0.40 $\pm$ 0.10 mag, and $\mu_0$ = 29.92 $\pm$ 0.29 mag.
$^c$Adopted from \citet{Sahu13} using $t_0$ = 2,455,712.5 $\pm$ 1.0, $E(B-V)$ = 0.035 mag, and $\mu_0$ = 29.62 $\pm$ 0.05 mag \citep{Vinko12}.
$^d$Calculated in this work, adopted parameters: $t_0$ = 2,456,447.8 $\pm$ 0.5, $E(B-V)$ = 0.09 $\pm$ 0.01 mag, $\mu_0$ = 31.10 $\pm$ 0.05 mag \citep{Freedman01,VanDyk14}.
\end{table*} 

\subsection{Analysis of the early bolometric light curve}\label{lc_bol}

In stripped envelope SNe, it has been observed that the shape of their light curves around the secondary peak looks more-or-less similar \citep[see e.g.][hereafter WJC15]{Wheeler15}. 
By the analysis of \citet{Arnett82}, this portion of the light curves can be utilized to extract
important physical parameters, such as the mass and the initial radius of the ejecta, the initial nickel mass, or the initial thermal and kinetic energy. Several authors have
employed this method to study stripped envelope SNe \citep[see e.g.][]{Clocchiatti97,Valenti08,Cano13,Wheeler15}. 

To estimate the values of the main explosion parameters for SN~2013df, we fitted a simple semi-analytic model to the bolometric light curve calculated from our early optical photometric 
measurements and from the UV data of {\it Swift}. We constructed this quasi-bolometric light curve using the method described in \citet{Marion14}.
To calculate the spectral energy distributions (SEDs) of SN~2013df, we converted the {\it BVRI} magnitudes to $F_{\lambda}$ fluxes using the calibration of 
\citet{Bessell98}. Since our {\it BVRI} measurements are not well-sampled around the secondary maximum, we calulated the missing values by interpolation using our {\it g'r'i'z'} photometry.
For the {\it Swift/UVOT} data we applied the latest zero-points and flux calibration by \citet{Breeveld11}.
The fluxes were dereddened using the Galactic reddening law parametrized by \citet{Fitzpatrick07} assuming $R_\rmn{V} = 3.1$ and adopting $E(B-V)=0.09$ mag (see Section \ref{obs}).
The quasi-bolometric light curve was derived by integrating the dereddened $F_{\lambda}$ values of the combined UV-optical SEDs over wavelength. 
The missing far-UV fluxes were estimated by assuming a linearly declining SED below 1928 \AA\ (the central wavelength of {\it Swift} UW2 filter) reaching zero flux at 1000 \AA.
The long-wavelength (near-and mid-infrared) contribution was estimated by fitting a Rayleigh-Jeans tail to the red end of the observed SEDs, and integrating it to infinity. 
Finally, the integrated fluxes were converted to luminosities using $D=16.6$ Mpc (see Section \ref{obs}).
The luminosity at the secondary maximum is around L$_\rmn{bol}$ $\sim$ 1.1 $\times$ 10$^{42}$ erg s$^{-1}$. We note that this value is approximately a factor of two lower than the one 
published by MG14; this difference may be due to mainly the different values of distance moduli discussed above.

The model we used incorporates recombination and was originally presented by \citet{Arnett89}; it has been recently extended by \citet{Nagy14} and \citet{Nagy16}, focusing on 
SNe II-P and IIb.
This extended model assumes a homologously expanding spherical ejecta, and pure Thomson scattering with opacities being constant in a given ionized layer. 
The model solves the photon diffusion equation in the ejecta, taking into 
account the energy release of the recombination processes, the radioactive heating of $^{56}$Ni and $^{56}$Co and the effect of gamma-ray leakage \citep{Chatzopoulos12}.

\begin{figure}
\begin{center}
\leavevmode
\includegraphics[width=.5\textwidth]{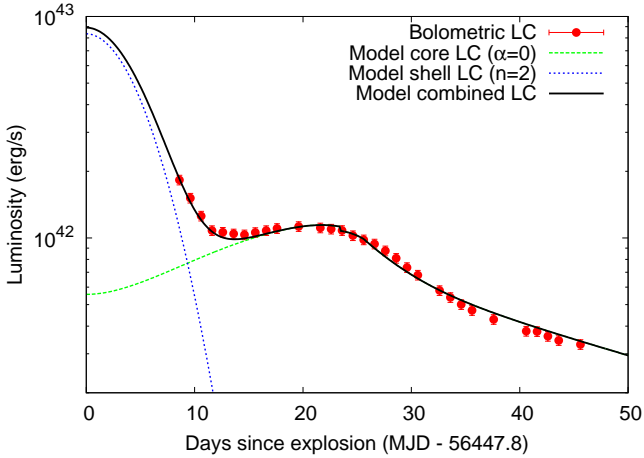}
\end{center}
\caption{Quasi-bolometric light curve of SN 2013df with the best-fitting semi-analytic model. See the details in the text.}
\label{fig:lcbol}
\end{figure}

Double-peaked light curves, as that of SN~2013df, are generally modeled by a two-component ejecta configuration \citep[see e.g.][]{Bersten12,Kumar13,Nagy16}. This two-component model 
contains an extended, low-mass, H-rich outer envelope and a more massive, denser, He-rich inner core. The contribution of the two different components is well separated in time, because 
the photon diffusion timescale is much lower in the outer region than in the core \citep[see][]{Kumar13}. At early phases, the radiation from the adiabatically cooling layer of the 
shock-heated outer envelope dominates the LC, while the second peak is powered by the radioactive decay of $^{56}$Ni and $^{56}$Co heating the inner core. The shape of the observed 
light curve can be modeled as the sum of these two processes.
We present the quasi-bolometric light curve of SN 2013df with the best-fitting semi-analytic model in Figure \ref{fig:lcbol}.

We used a constant density profile \citep[$\alpha = 0$ within the model of][]{Nagy14} for the inner region, while the outer H-rich shell had a power-law density distribution. 
The best fitting was obtained by using a density profile exponent of $n = 2.0 \pm 0.2$ ($\alpha$ and $n$ were fitted independently). In the outer shell the energy input from 
recombination is not significant because of the low mass and the adiabatic expansion of this region. 

Choosing proper values for the
optical opacities in the inner core as well as in the outer shell is a fundamentally important step in modelling the light curve with a semi-analytic model \citep[see][for a recent discussion of this issue]{Nagy16}. 
Based on our calculations, detailed below, we selected $\kappa \sim$ 0.4 cm$^2$ g$^{-1}$ and $\sim$ 0.2 cm$^2$ g$^{-1}$ for the outer H-rich envelope and the inner He-rich core, respectively. 
Setting $\kappa \sim 0.3$ cm$^2$ g$^{-1}$ for the outer envelope (which is closer to the expected mean opacity of a 
solar-like composition), the best-fit light curves are practically identical to those from the higher opacity models, resulting in very similar ejecta parameters from the fits.

To verify the reality of the chosen (constant) $\kappa$ values, we followed the method described in \citet{Morozova15}. We generated a presupernova model using 
MESA\footnote{http://mesa.sourceforge.net} \citep[Modules for Experiments in Stellar Astrophysics,][]{Paxton11,Paxton13} 1D stellar evolution code. First, we constructed a single star with 
an initial mass of $M_\rmn{in}$ = 18.5 M$_{\odot}$. Secondly, we removed the large part of the outer H-layer, getting a 7.5 M$_{\odot}$ star with a 1 M$_{\odot}$ H-layer as a synthetic 
SN IIb pre-explosion model.
This model was exported to the SNEC\footnote{http://stellarcollapse.org/snec} \citep[The SuperNova Explosion Code;][]{Morozova15} radiation-hydrodynamics code, which computes synthetic 
bolometric light curves following the expansion of an input presupernova model star. 
We found that the average opacities of those SNEC models that generate double-peaked Type IIb-like light curves are in good agreement with the ones we used during the analytic light curve 
modelling of SN~2013df. More details on the comparison between the opacities of SNEC models and other semi-analytic calculations, concerning both SNe II-P and IIb, can be found in \citet{Nagy16}. 

\begin{table*}
\begin{center}
\caption{Log of parameters derived from bolometric light curve modelling} 
\label{tab:lcbol_par}                     
\begin{tabular}{lcccl}          
\hline
\hline                      
Parameter &  Core (He-rich)  & Shell (mixed H-He) & Shell (pure H) & Remarks \\ 
& ($\kappa$ = 0.2 cm$^2$/g) &  ($\kappa$ = 0.3 cm$^2$/g)& ($\kappa$ = 0.4 cm$^2$/g) & ~ \\
\hline                          
$R_\rmn{0}$ (cm)& 7.4 $\times 10^{11}$ & 1.1 $\times 10^{13}$ & 1.2 $\times 10^{13}$ & Initial radius of the ejecta \\
$T_\rmn{rec}$ (K)& 10000$^a$ & -- & -- & Recombination temperature \\
$M_\rmn{ej}$ ($\mathrm{M_\odot}$)& 3.38$^b$  & 0.08$^b$ & 0.065$^b$ & Ejecta mass \\
$M_\rmn{Ni}$ ($\mathrm{M_\odot}$)& 0.043 & -- & -- & Initial nickel mass \\
$E_\rmn{Th}$ (foe)& 2.45 & 0.25 & 0.25 & Initial thermal energy \\
$E_\rmn{kin}$ (foe)& 2.7$^b$ & 0.2$^b$ & 0.2$^b$ & Initial kinetic energy \\
\hline
\hline
\end{tabular}
\end{center}
\smallskip
{\bf Notes.} $^a$ Adopted from \citet{Hatano99}. $^b$ Because of the $M_\rmn{{ej}} \kappa$ and $E_\rmn{kin} \kappa$ degeneracies, the $M_\rmn{ej}$ and $E_\rmn{kin}$ values 
given here are scaled to the opacities shown in the header. 
\end{table*}

The parameters derived from the modelling of the quasi-bolometric light curve of SN~2013df near peak are listed in Table \ref{tab:lcbol_par}. The values shown in the second and third columns pertain to 
the He-rich core and the outer shell assumed to consist of a mixture of H and He ($\kappa = 0.3$ cm$^2$ g$^{-1}$), respectively. We also show the results originating from the best-fit model using 
a pure H-shell ($\kappa = 0.4$ cm$^2$ g$^{-1}$). As mentioned above, these two cases result in very similar parameter sets.

In the following we discuss three important details. 
First, it is not possible to derive the independent values of the opacity, the ejected mass ($M_\rmn{ej}$) and the kinetic energy ($E_\rmn{kin}$) 
by this method. Instead, only the degenerate combinations of $M_\rmn{ej} \kappa$ or $E_\rmn{kin} \kappa$ can be constrained by the observations (see WJC15 for details). 
This needs to be taken into account if one compares the values for the ejecta masses and kinetic energies originating from various sources.
Thus, for example, the $M_\rmn{ej}$ and $E_\rmn{kin}$ values concerning the He-rich core need to be expressed as ($M_\rmn{ej} \kappa$)/(0.2 cm$^2$ g$^{-1}$) and ($E_\rmn{kin} \kappa$)/(0.2 cm$^2$ g$^{-1}$), 
respectively (as noted in Table \ref{tab:lcbol_par}). Using $\kappa =$ 0.2 cm$^2$ g$^{-1}$ for the optical opacity of the core, we get $M_\rmn{ej} \sim$3.4 $M_\rmn{\odot}$ and $E_\rmn{kin} 
\sim$2.7 $\times$ 10$^{51}$ erg. Varying the values of $\kappa$ in a range of 0.10 and 0.24 cm$^2$ g$^{-1}$ results in masses between 3.2$-$4.6 $\mathrm{M_\odot}$ and 
kinetic energies of $\sim 2.6 - 2.8 \times 10^{51}$ erg. 
For the best-fitting models the initial radius of the core, $R_\rmn{0, core}$, and the initial thermal energy, $E_\rmn{Th}$, also varied between 5$-$8 $\times 10^{11}$ cm and 2.3$-$2.6 $\times$ 10$^{51}$ erg, respectively.

Second, the $M_\rmn{ej}$ and $E_\rmn{kin}$ values concerning stripped envelope SNe can be very different, depending if they are determined from the modelling of the 
peak or the tail of the light curve (WJC15). This problem will be discussed in Section \ref{lc_late}.

Third, it is worth comparing our results to the previously published ones. MG14 also presented a light curve analysis based on the method of \citet{Arnett82}; however, their parameter set 
is quite different from ours. As discussed in Section \ref{lc_lc}, MG14 used a larger value for the distance than we did, which leads to systematic differences in the absolute magnitudes 
and luminosities. These differences obviously lead to different initial nickel masses ($M_\rmn{Ni} \sim$0.04 $\mathrm{M_\odot}$ vs. $\sim$0.10-0.13 $\mathrm{M_\odot}$ determined by 
MG14). 

There are also relatively large differences between our values and those of MG14 concerning the ejecta mass ($\sim$3.4 $M_\rmn{\odot}$ vs. 0.8$-$1.4 $M_\rmn{\odot}$) and the initial 
kinetic energy ($\sim$2.7 $\times$ 10$^{51}$ erg vs. 0.4$-$1.2 $\times$ 10$^{51}$ erg). We suggest that the main reason of these differences connects to the time scales.
Based on the model of \citet{Arnett82}, $M_\rmn{ej}$ and $E_\rmn{kin}$ can be determined from two peak parameters, the characteristic ejecta velocity, $v_\rmn{sc}$, and the rise 
time, $t_\rmn{rise}$ (see e.g. WJC15):
\begin{equation}
\label{eq:M_E}
\rmn{a)} \quad M_\rmn{ej} = \frac{1}{2} \frac{\beta c}{\kappa} v_\rmn{sc} t_\rmn{rise}^2, \quad \rmn{b)} \quad E_\rmn{kin} = \frac{3}{20} \frac{\beta c}{\kappa} v_\rmn{sc}^3 t_\rmn{rise}^2 ,
\end{equation}

\noindent where $\beta$ = 13.8 is an integration constant, $c$ is the speed of light. $t_\rmn{rise}$ is assumed to be equivalent with the mean 
light-curve time scale, $\tau_m$ = $\sqrt{2 \tau_d \tau_h}$, where $\tau_d$ is the diffusion time scale and $\tau_h = R_0 / v_\rmn{sc}$ is the hydrodynamic time scale, respectively.
Nevertheless, these assumptions of the Arnett model are valid only in that case when $R_0$ is negligible, which is not true for SNe having extended progenitors, i.e. for SNe II-P or 
eIIb. For SN~2013df we derived $R_0$ = 7.4 $\times$ 10$^{11}$ cm ($\sim$11 R$_{\odot}$) for the He-rich core. In this case, the contribution of the initial 
thermal energy ($E_\rmn{th}$, deposited by the shock) to the light curve is significant (see Table \ref{tab:lcbol_par}). One needs to take into account that
the shock-heated ejecta also radiates, and that modifies the light curve shape considerably. 
Recombination is also a major process that can significantly affect the measured peak time of the light curve \citep{Arnett89,Nagy16}.
All of these details must be considered in modelling the light curves of Type II SNe.   
Otherwise, the naive estimation of $t_\rmn{rise}$ = $\tau_m$ may lead to the significant underestimation of $M_\rmn{ej}$ and $E_\rmn{kin}$.
For example, if we calculate $\tau_m$ from the best-fit model parameters listed in Table \ref{tab:lcbol_par}, we get 27.5 days, which is much larger than the observed rise time estimated to be $\sim$22 days.

We also note that there is a typo in \citet{Arnett82} where the relation between the photospheric velocity, the kinetic energy and the ejecta mass is given as 
$v_\rmn{phot}^2 = 3/5(2 E_\rmn{kin} / M_\rmn{ej})$, as recently pointed out by WJC15.  
This typo was corrected in \citet{Arnett96} where the proper relation was given as $v_\rmn{phot}^2 = 5/3(2 E_\rmn{kin} / M_\rmn{ej})$, 
but this error has propagated in the literature, for example into \citet{Valenti08} whose formulae were applied by MG14.

Our two-component model allowed us to extract information on the parameters of the outer shell including the initial radius of the extended H-rich material, $R_\rmn{0, shell}$; however, we note that determination of this parameter is very uncertain, 
especially, if the light curve is not well sampled in the initial declining phase. Thus, we consider the value of $R_\rmn{0, shell}$ $\sim$ 1.1$-$1.2 $\times$ 10$^{13}$ cm ($\sim$160$-$170 R$_{\odot}$) as a lower limit.
This conclusion does not contradict the results of VD14 who determined $R_\rmn{eff}$ = 545 $\pm$ 65 $R_{\odot}$ as the effective radius of the progenitor based on HST images obtained 14 years before explosion.
Since the model applied by MG14 can not be used for fitting the very early part of the bolometric light curve (i.e. the epochs before the secondary maximum), they could not estimate the shell parameters this way. Instead, 
they used the formulae of \citet{NP14} to constrain the mass and the radius of the shell. As they noted, this method also produces very uncertain results if there are no observed data in the very early phases of the light curve, or only in one filter 
(which is true for SN~2013df). Therefore, they also defined the value of the radius of the extended material of $R_{ext} \sim$64$-$169 $R_{\odot}$, which should be considered as a lower limit, while lacking the observed peak luminosity of the shock breakout. 

\subsection{Explosion parameters from late-time photometry}\label{lc_late}

Based on the method of \citet{Arnett82}, the properties of the late-time light curves can be predicted from the physical parameters inferred from the early part of the light curves of 
stripped envelope SNe.
The observed late-time light curves may vary substantially for SN Ib, Ic and IIb despite their similar early-time light curves \citep[see][]{Clocchiatti97,Wheeler15}.
The differences of the late-time light curve behavior are indicative of heterogeneous explosion kinematics and progenitor masses concerning this type of stellar explosions.
WJC15 suggested that there may be a discrepancy between the fitted tail decay parameters and those obtained from the fit to the rising part of the light curve.
That suggest incomplete physical modelling of the whole light curve.

Here, we address the issue of deriving physical parameters from the light-curve tail of SN~2013df, and compare late-time results to those obtained from the peak properties.
We used three data sets, which all consist of measurements obtained after +50d, to construct the late-time light curve. Our unfiltered ROTSE observations cover the epochs between 
+50 and +82 days, and also contain a single measurement obtained at +168d (note that some of these data have quite high uncertainties). Additionally, we adopted 
the late-time $r'$-band data of MG14 that extend up to +252d. 
These late-time data show significant epoch-to-epoch variations around an exponential decline of $\sim$0.019 mag day$^{-1}$.
We also included one R-band and three $r'$-band photometric points (obtained at +199, +262, +298, and +348 days, respectively) published by \citet{Maeda15}; 
we converted the R-band magnitude belonging to +199d into an $r'$-band value using the calibration by \citet{Jordi06}.
Following the description of \citet{Lyman14} and WJC15, we neglected the bolometric corrections and assumed that these late-time photometric values 
are proxies of the bolometric luminosities to within a constant scaling factor.
As earlier, we used $t_0$ = 2,456,447.8 $\pm$ 0.5 JD and $\mu_0$= 31.10 $\pm$ 0.05 mag while deriving the luminosities.

%
The late-time light curve of stripped envelope SNe is thought be powered by the energy input from the decay of $^{56}$Ni and $^{56}$Co, mainly via the thermalization of the emerging $\gamma$-ray photons trapped in the expanding material
\citep[see e.g.][as well as WJC15, and references therein]{Cappellaro97,Vinko04}. Depending on the density of the ejecta, 
part of the high-energy photons leak out from the ejecta; the degree of $\gamma$-ray leakage is indicated by the steepness of the late-time light curve.
A fraction of the decay energy leads to the production of positrons, which are slowed down and annihilated by the ejecta material.

The rate of energy production in gamma-rays can be given as
\begin{equation}
E_{\gamma} = M_\rmn{Ni}\ \left[\epsilon_\rmn{Ni}\ e^{-t/\tau_\rmn{Ni}}\ +\ \epsilon_\rmn{Co}\ \left(e^{-t/\tau_\rmn{Co}}\ -\ e^{-t/\tau_\rmn{Ni}}\right)\right]\ ,
\end{equation}

\noindent where $M_\rmn{Ni}$ is the mass of the synthesized nickel, $\epsilon_\rmn{Ni}$ = 3.9 $\times$ 10$^{10}$ erg s$^{-1}$ g$^{-1}$ and $\epsilon_\rmn{Co}$ = 6.8 $\times$ 10$^9$ erg s$^{-1}$ g$^{-1}$ 
are the energy generation rates of $^{56}$Ni and $^{56}$Co, respectively \citep[see e.g.][]{Nadyozhin94,Arnett96}, while  $\tau_\rmn{Ni}$ = 8.8 d and $\tau_\rmn{Co}$ = 111.3 d are the decay times of $^{56}$Ni and $^{56}$Co, respectively.
The total radiated luminosity resulting from the gamma-ray trapping per seconds can be approximated as
\begin{equation}
L_\rmn{\gamma} = E_{\gamma}\ D_{\gamma}\ = E_{\gamma}\ \left(1\ -\ e^{- (T_0/t)^2}\right) ,
\end{equation}

\noindent where $D_\gamma$ is the gamma-ray deposition function and $T_\rmn{0}$ is defined as the characteristic timescale for the leakage of $\gamma$-rays.

Following \citet{Branch16}, we describe the rate of energy production due to positrons as \begin{equation}
E_{+} = M_\rmn{Ni}\ \left[\left(\epsilon_\rmn{kin}\ +\  \epsilon_\rmn{an}\ D_{\gamma}\right)\ \left(e^{-t/\tau_\rmn{Co}}\ -\ e^{-t/\tau_\rmn{Ni}}\right)\right]\ ,
\end{equation}

\noindent where $\epsilon_\rmn{kin}$ = 2.18 $\times$ 10$^8$ erg s$^{-1}$ g$^{-1}$ and $\epsilon_\rmn{an}$ = 3.63 $\times$ 10$^8$ erg s$^{-1}$ g$^{-1}$ are the energy generation rates of the thermalization 
and the annihilation of positrons, respectively \citep[calculated from][]{Nadyozhin94}. Using a similar approximation for positron trapping as for the gamma-rays, 
the total radiated power of positrons is given by the following formula:  
\begin{equation}
L_\rmn{+} = E_{+}\ D_{+}\ = E_{+}\ \left(1\ -\ e^{- (T_\rmn{+}/t)^2}\right) ,
\end{equation}

\noindent where $T_\rmn{+}$ is defined as the characteristic timescale for the leakage of positrons.
Note that the positron deposition involves two deposition functions. One, $D_+$, is intrinsic to the transport and trapping of the positrons and treated here as if there were an 
effective positron transport opacity. The other is the deposition function that also applies to gamma-rays, $D_\gamma$. The annihilation term involves a product of both deposition 
functions since the positrons must be stopped before they can annhilate, and the resulting gamma-rays are then subject to (approximately) the same deposition function as the gamma-rays 
produced directly in the decay.

The total late-time luminosity thus becomes the sum of the gamma-ray and positron energy production,
\begin{equation}
\label{eq:L}
L = L_\rmn{\gamma}\ +\ L_\rmn{+}\ .
\end{equation}

\begin{figure}
\begin{center}
\leavevmode
\includegraphics[width=.5\textwidth]{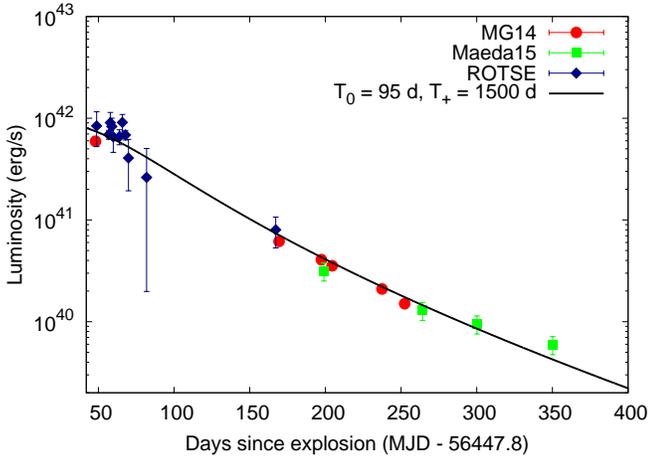}
\caption{The best fit model curve described by Eq. (\ref{eq:L}) to the late-time light curve of SN~2013df, which was calculated from our ROTSE data (diamonds), 
as well as from the late-time $r'$-band measurements of \citet[][MG14, with circles]{MG14} and \citet[][rectangles]{Maeda15}. See details in the text.}
\label{fig:tail_fit}
\end{center}
\end{figure}

While the deposition of positrons is often neglected or treated crudely (see e.g. WJC15 and references therein), we tried to handle them explicitly to examine 
their effect on the determination of the explosion parameters \citep[as was done with various approximations by e.g.][]{Cappellaro97,Vinko04,Valenti08}.
First, we assumed that the late-time light curve can be described using only one characteristic timescale, which yielded $T_0 = T_+ = 135 \pm 5$ days.
While this value may be realistic for the timescale of $\gamma$-ray leakage, it is too short for the timescale of positron leakage.
We thus adopted different timescales for the gamma-ray and positron depositions. Following \citet{Clocchiatti97}, we specified these timescales assuming 
constant density as
\begin{equation}
\label{eq:T0}
\rmn{a)} \quad T_0 = \sqrt{\frac{3\ \kappa_{\gamma}\ M_\rmn{ej}}{4\ \pi\ v_\rmn{phot}^2}},\quad \mathrm{and} \quad \rmn{b)} \quad T_\rmn{+} = \sqrt{\frac{3\ \kappa_{+}\ M_\rmn{ej}}{4\ \pi\ v_\rmn{phot}^2}}\ ,
\end{equation}

\noindent where $v_\rmn{phot}$ is the photospheric velocity at the peak, $\kappa_{\gamma} \approx 0.027 - 0.030$ cm$^2$ g$^{-1}$ \citep{Colgate80} and 
$\kappa_{+} \approx 7 - 10$ cm$^2$ g$^{-1}$ are the typical opacities of gamma-rays and positrons, respectively \citep{Colgate80,Milne99,Penney14}. 
We note that the value of $\kappa_{+}$ is quite uncertain. In all the cited papers, the transport of positrons and gamma-rays were modeled in SN Ia atmospheres; the results, based on 
Monte-Carlo simulations, imply that not only the degree of ionization but also the strength of the magnetic field need to be taken into account to give the exact value of $\kappa_{+}$.

In the lack of other information concerning SN IIb atmospheres, we chose $\kappa_{+}$ = 7 cm$^2$ g$^{-1}$ and $\kappa_{\gamma}$ = 0.028 cm$^2$ g$^{-1}$ for our further investigations. 
Using these values, the relations given in Eq. \ref{eq:T0} imply that $T_+/T_0 = \sqrt{\kappa_+/\kappa_\gamma} \approx 15.8$. 
Taking this condition into account, the second fit to the late-time light curve yielded $T_0 = 95 \pm 1$ days and $T_\rmn{+} = 1500 \pm 15.8$ days. 
This large value of $T_\rmn{+}$ seems to be consistent with the commonly accepted hypothesis that positrons are almost completely trapped in the ejecta, but a fraction of them might escape 
\citep[see e.g.][]{Clocchiatti97}. The best fit model to the late-time light curve, described by Eq. (\ref{eq:L}) and applying the boundary condition for $T_\rmn{+}$ discussed above, 
is shown in Figure \ref{fig:tail_fit}. The smaller value of $T_0$ is closer to the values predicted from the peak by WJC15.

Using the characteristic timescales determined from the second fit, and $v_\rmn{phot}$ = 9000 km s$^{-1}$ for the photospheric velocity at the secondary maximum (see Table \ref{tab:synpar}), 
we derived the ejecta mass of $M_\rmn{ej}$ = 3.75$-$4.20 $\mathrm{M_\odot}$ from Eq. (\ref{eq:T0}a); the lowest and the highest value of $M_\rmn{ej}$ belongs to $\kappa_{\gamma}$ = 0.030 and 
$\kappa_{\gamma}$ = 0.027, respectively. 
From the values of $T_0$ and $M_\rmn{ej}$ determined from the second fit, the kinetic energy can be calculated as
\begin{equation}
\label{eq:Ekin}
E_\rmn{kin} = (9 \kappa_{\gamma} M_\rmn{ej}^2)/(40 \pi T_0^2) ,
\end{equation}

\noindent which gives $E_\rmn{kin}$ = 1.56$-$2.30 $\times$ 10$^{51}$ erg. 

The values of the ejecta mass and the kinetic energy originating from the analysis of the late-time light curve employing separate timescales for the gamma-ray and positron deposition 
agree well with those determined from the modelling of the early-time bolometric light curve (see Section \ref{lc_bol}).
If we had used the $T_0$ value from the first fit of the late-time light curve (where we assumed the same timescale for the leakage of $\gamma$-rays and positrons), 
we would get a factor of $\sim 2$ larger ejecta mass. This implies that the correct handling of the positron deposition functions may play an important role in the determination of the 
explosion parameters via the analysis of light-curve tails of stripped envelope SNe.

WJC15 found significant inconsistencies in the ejecta masses (as well as in the kinetic energies) originating from the analysis of the peak and of the tail of the light curves. 
They suggested that those inconsistencies could be resolved by assuming very small optical opacities ($\kappa \sim$0.01$-$0.1 g cm$^{-2}$) in the ejecta around the maximum 
\citep[the very low values of $\kappa$ in the ejecta of stripped envelope SNe were also assumed by e.g.][]{Ensman88,Kleiser14}.
In the case of SN~2013df, there are two steps seem to be important in solving this problem. First, we developed a model, in which $M_\rmn{ej}$ and $E_\rmn{kin}$ are input parameters, thus 
they can be directly determined from the fitting. This allowed us to avoid to calculate these parameters from the rise time to the secondary peak, which is usually assumed to have an exact 
physical meaning ($t_\rmn{rise}$ = $\tau_m$ = $\sqrt{2 \tau_d \tau_h}$), but which is only true in the cases of compact progenitors. Second, we tried to handle explicitly the effects 
of the deposition of gamma-rays and positrons during the analysis of the light-curve tail.

Nevertheless, there are several other theoretical 
and observational aspects that need to be taken into account, and the findings concerning SN 2013df do not necessarily help to answer the open questions concerning other SN IIb, or, 
even more, SN Ib/c. Both the peak and tail models contain simplifications and uncertainties, whose effects on the parameter sets are not exactly 
known. Moreover, the photometric errors or the lack of multicolor observations concerning the late-time data, the uncertainties of some observational parameters (especially those of 
$t_0$ and $v_\rmn{phot}$), as well as the systematic uncertainties of $\mu_0$ may also significantly affect the inferred explosion parameters.
As WJC15 also noted, these problems only can be solved via careful analyses of homogeneous, high-quality multicolor data that cover both the peak and the tail of 
the light curves well.

\subsection{Signs of CSM interaction from late-time {\it Spitzer} photometry}\label{mir}

In addition to the analysis based on our optical spectroscopic and photometric data, we also studied the late-time mid-IR data of SN~2013df.
Up to now, SN~2011dh was the only SN IIb with published late-time mid-IR fluxes \citep{Helou13,Ergon14,Ergon15}. Very recently, \citet{Tinyanont16} published the results of 
the SPIRITS (SPitzer InfraRed Intensive Transients Survey, PI: M. Kasliwal) program, during that they monitored several nearby galaxies (and transients in them) in the near past 
with the {\it Spitzer Space Telescope}. This dataset includes measurements about the host galaxy of SN~2013df obtained at five epochs between +264 and +824 days.

\begin{figure*}
\begin{center}
\includegraphics[width=.95\textwidth]{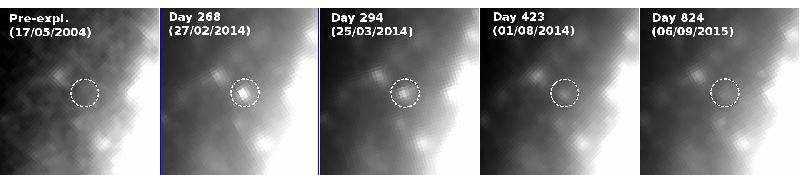}
\includegraphics[width=.95\textwidth]{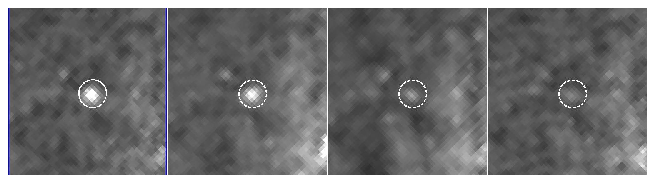}
\caption{The mid-IR evolution of SN~2013df on Spitzer 4.5 $\mu$m images (top: original PBCD images, bottom: background-subtracted images).}
\label{fig:mir}
\end{center}
\end{figure*}

Although the detailed analysis of the {\it Spitzer} data of SN~2013df is beyond the scope of this paper, the study of the mid-IR light curves offers a good chance to look into the 
late-time evolution of the SN. 
As it can be seen in Figure \ref{fig:mir}, the object is clearly detectable in mid-IR even more than two years after explosion (for the better visibility, we also present 
background-subtracted images created with HOTPANTS\footnote{http://www.astro.washington.edu/users/becker/hotpants.html; developed by A. Becker}, using a set of pre-explosion IRAC images 
obtained in 2004 by Fazio et al. as templates).

\begin{figure*}
\begin{center}
\includegraphics[width=.45\textwidth]{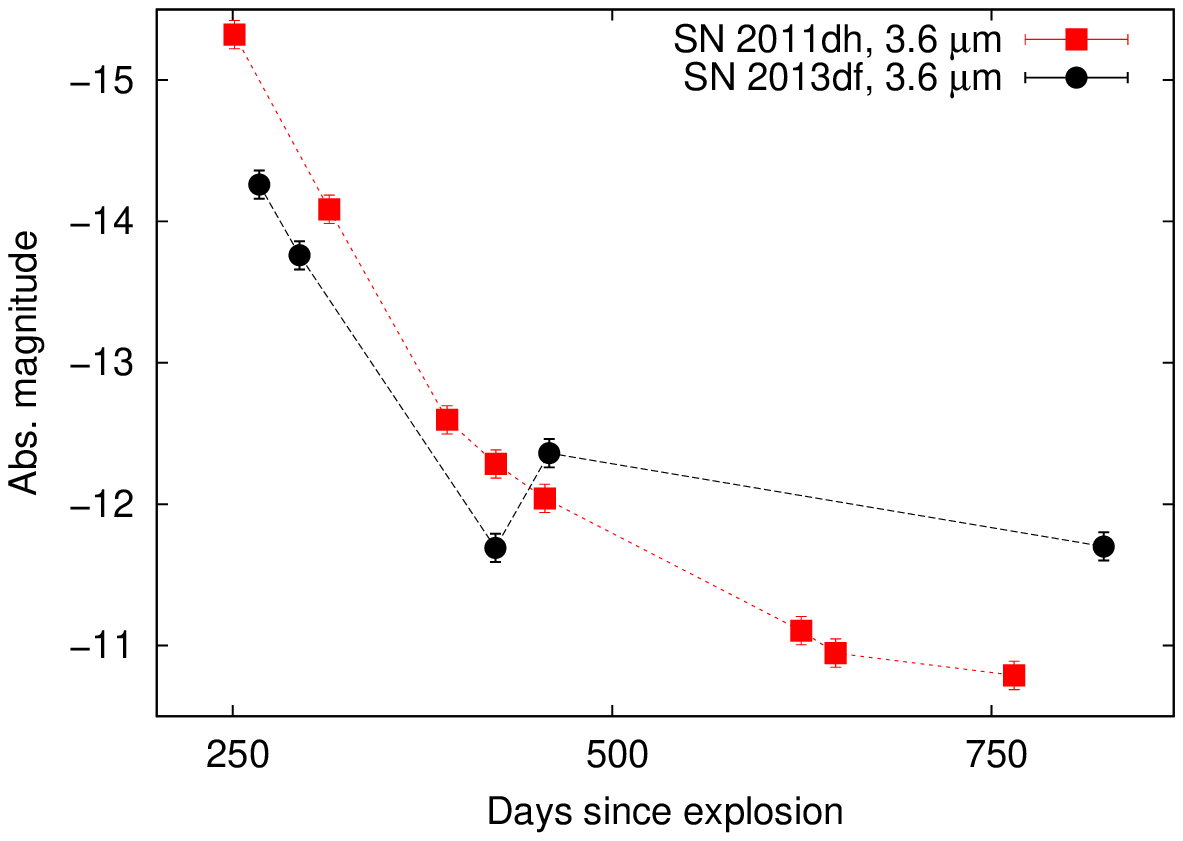} \hspace{5mm}
\includegraphics[width=.45\textwidth]{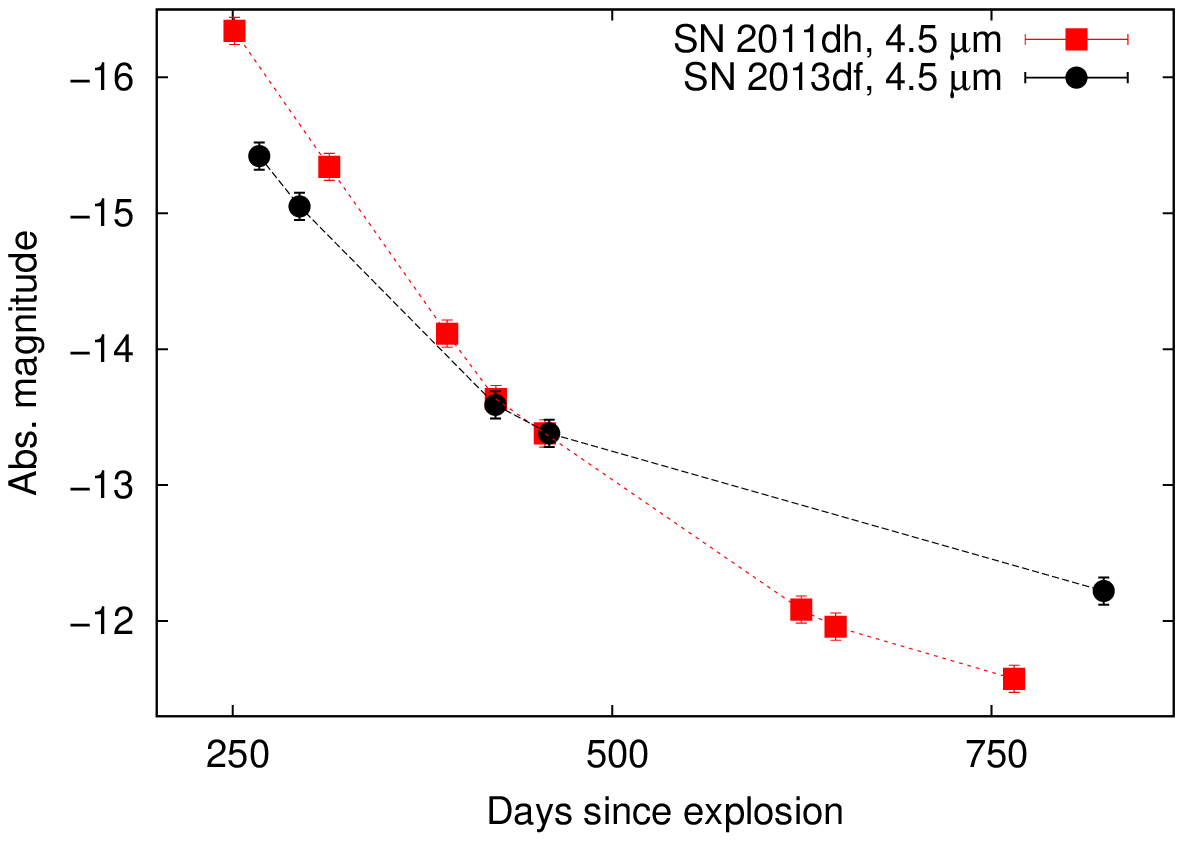}
\caption{Late-time 3.6 (left panel) and 4.5 micron (right panel) light curves of SNe~2011dh \citep[][filled rectangles]{Ergon15} and 2013df \citep[][filled circles]{Tinyanont16}. 
Absolute magnitudes were calculated using the distances of D = 8.4 Mpc and 16.6 Mpc, respectively.}
\label{fig:mir_comp}
\end{center}
\end{figure*}

Figure \ref{fig:mir_comp} shows the comparison of late-time mid-IR light curves of SN~2013df \citep{Tinyanont16} and those of SN~2011dh \citep{Ergon15}. 
We scaled the {\it Spitzer} fluxes of SN~2011dh to the distance of SN~2013df (D = 16.6 Mpc), adopting D = 8.4 Mpc for the 
distance of SN~2011dh \citep{Marion14}. 
While the mid-IR evolution of the two objects seem to be similar between $\sim$250d and 350d (in this period, SN~2011dh seems to be even brighter than SN~2013df),
none of the 3.6 and 4.5 $\mu$m light curves of SN~2013df follow the drop of those of SN~2011dh after +350d (moreover, SN~2013df seems to show a rebrightening at 3.6 $\mu$m).

\citet{Helou13} found that the late-time mid-IR behavior of SN~2011dh cannot be explained by a simple thermal echo model but only with additional dust heating or line emission mechanisms.
Therefore, the source of the more slower decreasing late-time mid-IR excess of SN~2013df must represent an even larger amount of dust formed in the ejecta 
\citep[see e.g.][]{Kozasa09,Nozawa10} or formed/heated during the interaction of the ejecta and the sorrunding circumstellar material \citep[as has been found in SNe IIn, see][]{Fox11,Fox13}.
Both processes can take place in SN IIb some hundreds of days after explosion \citep[see][for a review]{Gall11}.

This finding is consistent with the results of \citet{Maeda15} who found signs of strong CSM interaction (emergence of broad and flat-topped H$\alpha$ and He\,\begin{small}I\end{small} 
emission lines), beginning at $\sim$1 yr after explosion, in the late-time optical spectra of SN~2013df. Very similar effect was found in the case of the well-known interacting SN IIb 1993J 
\citep{Matheson00,Matheson01}, while there was only a marginal detection of a broad H$\alpha$ line in the late spectra of SN~2011dh \citep[see][]{Shivvers13,Maeda15}.
Late-time radio and X-ray behavior of SN~2013df is also very similar to that of SN~1993J, indicating also ongoing CSM interaction \citep[see][and references therein]{Kamble16}.

The late-time CSM interaction may also affect the optical light-curve tail of SN~2013df. As seen in Figure \ref{fig:tail_fit}, the last point at $\sim$350d is well above the best-fit model. 
This effect can be explained with the emergence of the strong H$\alpha$ emission, which seems to appear as an excess in the $r'$-band flux.

\section{Conclusions}\label{conc}

According to previous results, we found SN~2013df to be an extended Type IIb supernova, showing a close spectral similarity with SN 1993J. While the light curve evolutions of the two 
SNe are also similar, SN~2013df seems to be less luminuous than SN~1993J (but there are relatively large uncertainties concerning the distance of the host galaxy of SN~2013df).
The application of the SYNAPPS spectral synthesis code allowed us to determine some important physical parameters of the ejecta: we found a continuously decreasing photosperic 
temperature from 8000 K to 5800 K, and a decreasing photospheric velocity from $\sim$14\,000 km s$^{-1}$ to 8000 km s$^{-1}$ between +9 and +34 days.
While the temperatures are consistent with the ones published by MG14 and BA15, the photospheric velocities are significantly larger than those previously reported concerning SN~2013df.
The exact determination of the photosperic velocity at the secondary peak is specifically important, because it is used as an input parameter in the calculation of the ejecta mass and the 
initial kinetic energy via fitting the light-curve tail.

Our most important result concerning the spectral analysis is the identification of ``high-velocity'' He\,\begin{small}I\end{small} features in the early spectra. At +9d and +12d, the 
He\,\begin{small}I\end{small} $\lambda 5876$ profile could be fit with a single He\,\begin{small}I\end{small} line at $v \sim 14,000$ km s$^{-1}$. A lower velocity component at 
$v \sim 7,800$ km s$^{-1}$ appears after +19d. Similar but less conspicuous HV feature are also seen in the He I $\lambda 6678$ line. 
Similar spectral properties are seen in the cases of SNe 1993J and 2011fu, both of which belong also to SN eIIb. On the other hand, the spectra of several other SN IIb, which are
thought to emerge from more compact progenitors, do not show this effect.
Moreover, while most SN IIb can be characterized by a clear separation of H and He lines in velocity space ($\sim$4-5\,000 km s$^{-1}$), such a separation is not seen in the case of either SNe~2013df, 
1993J or 2011fu, as well as in the light echo spectra of Cas A (which all thought to emerge from extended progenitors). These findings indicate that the early ``high-velocity'' 
He features might form in the low-mass H-rich envelope on top of the denser, He-rich core. The evolution of the He lines can be described by the inward motion of the photosphere from the 
high-velocity outer envelope to the more slowly-expanding core.
The presence of this effect may depend on the degree of the mixing of H and He layers, which may be more significant in extended supergiant stars than in more compact progenitors 
(e.g. Wolf-Rayet stars). This is a possible explanation why we only see this non-separation of H$\alpha$ and He\,\begin{small}I\end{small} $\lambda 5876$ velocities in the cases of
SN eIIb. 
Nevertheless, further studies are necessary to reveal the currently unknown details of this effect, which may require even more sophisticated tools e.g. Monte Carlo 
radiative transfer codes.

We fitted a semi-analytic model to the early bolometric light curve of SN~2013df. The derived explosion parameters are consistent with those of other SN~IIb (taking into account the 
uncertainties concerning the distances). 
We found a lower limit of the initial radius of the ejecta of $\sim$160 R$_{\odot}$, which is consistent with the result of MG14, and does not contradict the progenitor radius 
$R_\rmn{eff} \sim$545 R$_{\odot}$ estimated by VD14.
The analysis of the late-time light curve decline, taking into account the leakage of both the $\gamma$-rays and positrons originating from the 
$^{56}$Co-decay, resulted in an ejecta mass (3.7$-$4.2 $M_\rmn{\odot}$) and kinetic energy (1.6$-$2.3 $\times$ 10$^{51}$ erg) that are in good agreement with the same parameters estimated 
from fitting the secondary peak of the light curve (3.2$-$4.6 $M_\rmn{\odot}$ and 2.6$-$2.8 $\times$ 10$^{51}$ erg, respectively). 
This finding seems to contradict the general discrepancy, found by WJC15, concerning the ejecta masses end kinetic energies inferred from the early- and late-time light curves of stripped envelope SNe. 
We suggest that there are two key steps in our analysis that allowed us to find seemingly consistent solutions for 
$M_\rmn{ej}$ and $E_\rmn{kin}$ in the case of SN~2013df: first, the careful handling of time scales and relations originating from Arnett's model during 
the analysis of the light-curve peak, and, second, the explicit handling of positron deposition during the light-curve tail fitting. 
On the other hand, it should be kept in mind that the models we applied for the spectrum synthesis
and for the light curve fitting have limitations and caveats. For example, beside the effect of the 
constant opacity, as discussed in \citet{Nagy16}, the light curve model is clearly affected by
the assumption of the centrally peaked distribution of $^{56}$Ni. There are more complex 
modeling codes available, like SNEC \citep{Morozova15} or TARDIS \citep{Kerzendorf14},
which, when fully developed, will be promising future improvements to the simple calculations
presented in this paper.
Nevertheless, there are a lot of open questions concerning the 
determination of the explosion parameters of stripped envelope SNe; in order to find the answers to them, well-sampled datasets need to be obtained, which should be carefully analysed 
applying either analytic or hydrodynamic codes. 

We also studied late-time mid-IR data obtained with {\it Spitzer}. Both the 3.6 and 4.5 $\mu$m light curves of SN~2013df show an excess to those of SN~2011dh between $\sim$+350 and +825 days.
This indicates circumstellar interaction starting $\sim$1 year after explosion, in accordance with previously published optical, X-ray, and radio data.

\section*{Acknowledgments} 

This work has been supported by the Hungarian Scientific Research Fund (OTKA) Grants NN107637, K104607, K83790, and K113117.
TS is supported by the OTKA Postdoctoral Fellowship PD112325.
JCW’s Supernova group at the UT Austin is supported by NSF Grant AST 11-09881 grant. 
JMS is supported by an NSF Astronomy and Astrophysics Postdoctoral Fellowship under award AST-1302771.
KS and AP are supported by the 'Lend\"{u}let-2009' Young Researchers Program and the LP2012-31 grant of the Hungarian Academy of Sciences, respectively;
KS is also supported by the ESA PECS Contract No. 4000110889/14/NL/NDe.

The Hobby-Eberly Telescope (HET) is a joint project of the University of Texas at Austin, the Pennsylvania State University, 
Stanford University, Ludwig-Maximilians-Universit\"at M\"unchen, and Georg-August-Universit\"at G\"ottingen. 
The HET is named in honor of its principal benefactors, William P. Hobby and Robert E. Eberly.
The Marcario Low Resolution Spectrograph is named for Mike Marcario of High Lonesome Optics 
who fabricated several optics for the instrument but died before its completion. 
The LRS is a joint project of the Hobby-Eberly Telescope partnership and the Instituto de Astronom\'ia 
de la Universidad Nacional Aut\'onoma de M\'exico.
We acknowledge the thorough work of the HET resident astronomers, Matthew Shetrone, Stephen Odewahn, John Caldwell
and Sergey Rostopchin during the acquisition of the spectra.

ROTSE IIIb telescope operation and data analysis is supported by NASA grant NNX10A196H (P.I. Kehoe).

This research has made use of the NASA/IPAC Extragalactic Database (NED) which is operated by the Jet Propulsion Laboratory, California Institute of Technology, under contract 
with the National Aeronautics and Space Administration. We acknowledge the availability of NASA ADS services.

\appendix

\section[]{Photometric results}

\begin{table*}
\begin{center}
\caption{{\it BVRI} (Konkoly Observatory, Hungary) magnitudes of SN~2013df}
\label{tab:mag1}
\begin{tabular}{@{}cccccc@{}}
\hline
\hline
JD$^a$ & Phase$^b$ & B & V & R & I \\
~ & (days) & (mag) & (mag) & (mag) & (mag) \\
\hline
6456.4 & 8.6 & 14.78(.07) & 14.51(.04) & 14.41(.04) & 14.19(.02) \\ 
6457.4 & 9.6 & 15.05(.07) & 14.71(.02) & 14.56(.04) & 14.33(.01) \\ 
6458.4 & 10.6 & 15.20(.10) & 14.86(.03) & 14.68(.04) & 14.47(.01) \\ 
6459.4 & 11.6 & 15.37(.02) & 14.91(.14) & 14.77(.05) & 14.46(.01) \\ 
6460.4 & 12.6 & 15.41(.03) & 14.92(.03) & 14.67(.07) & 14.49(.01) \\ 
6461.4 & 13.6 & 15.43(.03) & 14.88(.04) & 14.64(.04) & 14.45(.02) \\ 
6462.4 & 14.6 & 15.39(.18) & 14.84(.01) & 14.59(.05) & 14.41(.01) \\ 
6463.4 & 15.6 & ... & 14.77(.01) & 14.50(.02) & 14.36(.02) \\ 
6464.4 & 16.6 & 15.38(.08) & 14.71(.02) & 14.46(.06) & 14.32(.01) \\ 
6465.4 & 17.6 & 15.39(.01) & 14.66(.10) & 14.40(.02) & 14.27(.01) \\ 
6467.4 & 19.6 & 15.40(.01) & 14.57(.01) & 14.30(.06) & 14.19(.01) \\ 
6474.4 & 26.6 & 15.56(.16) & 14.71(.01) & 14.35(.03) & 14.16(.12) \\ 
6475.4 & 27.6 & 15.79(.11) & 14.76(.02) & 14.38(.06) & 14.21(.01) \\ 
6476.4 & 28.6 & 15.89(.02) & 14.84(.02) & 14.42(.04) & 14.25(.01) \\ 
6477.4 & 29.6 & 16.12(.03) & 14.95(.01) & 14.50(.06) & 14.30(.03) \\ 
6478.4 & 30.6 & 16.33(.02) & 15.08(.02) & 14.57(.02) & 14.35(.02) \\ 
6480.4 & 32.6 & 16.50(.01) & 15.27(.01) & 14.72(.07) & 14.51(.01) \\ 
6481.4 & 33.6 & 16.68(.04) & 15.39(.03) & 14.79(.03) & 14.55(.02) \\ 
6482.4 & 34.6 & 16.87(.01) & 15.49(.02) & 14.83(.06) & 14.59(.01) \\ 
6483.4 & 35.6 & 16.83(.01) & 15.57(.01) & 14.91(.05) & 14.66(.03) \\ 
6485.4 & 37.6 & 16.99(.02) & 15.66(.01) & 15.02(.06) & 14.75(.01) \\ 
6488.4 & 40.6 & 17.21(.13) & 15.83(.01) & 15.15(.05) & 14.87(.01) \\ 
6489.4 & 41.6 & 17.19(.08) & 15.83(.02) & 15.18(.07) & 14.85(.01) \\ 
6490.4 & 42.6 & 17.17(.09) & 15.91(.01) & 15.23(.06) & 14.91(.02) \\ 
6491.4 & 43.6 & 17.44(.10) & 15.92(.06) & 15.27(.06) & 14.94(.01) \\ 
6493.4 & 45.6 & 17.10(.11) & 15.99(.04) & 15.31(.04) & 15.00(.01) \\ 
\hline
\end{tabular}
\end{center}
\smallskip
{\bf Notes.} $^{(a)}$JD$-$2,450,000. $^{(b)}$ With respect to $t_{0}$ = 2,456,447.8 $\pm$ 0.5 JD. Errors are given in parentheses.
\end{table*}

\begin{table*}
\begin{center}
\caption{{\it g'r'i'z'} (Baja Observatory, Hungary) magnitudes of SN~2013df}
\label{tab:mag2}
\begin{tabular}{@{}cccccc@{}}
\hline
\hline
JD$^a$ & Phase$^b$ & g' & {\it r'} & i' & z' \\
~ & (days) & (mag) & (mag) & (mag) & (mag) \\
\hline
6455.8 & 8.0 & 14.59(.03) & 14.52(.02) & 14.49(.03) & 14.58(.04) \\
6456.8 & 9.0 & 14.86(.03) & 14.68(.03) & 14.64(.03) & 14.69(.04) \\
6457.9 & 10.1 & 14.98(.05) & 14.76(.04) & 14.82(.05) & 14.92(.07) \\
6458.8 & 11.0 & 15.09(.05) & 14.83(.04) & 14.82(.05) & 14.72(.05) \\
6459.9 & 12.1 & 15.12(.03) & 14.81(.02) & 14.81(.03) & 14.84(.04) \\
6460.9 & 13.1 & 15.06(.03) & 14.76(.03) & 14.80(.03) & 14.84(.04) \\
6461.9 & 14.1 & 15.06(.03) & 14.72(.03) & 14.71(.03) & 14.71(.04) \\
6462.9 & 15.1 & 15.01(.03) & 14.60(.03) & 14.64(.04) & 14.75(.04) \\
6463.9 & 16.1 & 14.97(.04) & 14.60(.03) & 14.63(.04) & 14.69(.04) \\
6464.9 & 17.1 & 15.00(.04) & 14.55(.04) & 14.62(.04) & 14.73(.05) \\
6466.9 & 19.1 & 14.84(.04) & 14.42(.03) & 14.53(.04) & 14.54(.04) \\
6468.9 & 21.1 & 14.91(.04) & 14.33(.03) & 14.49(.04) & 14.69(.04) \\
6469.9 & 22.1 & 14.89(.03) & 14.38(.02) & 14.46(.03) & 14.60(.04) \\
6470.9 & 23.1 & 14.93(.03) & 14.35(.02) & 14.39(.03) & 14.51(.03) \\
6471.9 & 24.1 & 14.98(.04) & 14.39(.03) & 14.42(.03) & 14.55(.04) \\
6472.9 & 25.1 & 15.05(.03) & 14.40(.02) & 14.49(.03) & 14.55(.03) \\
6473.9 & 26.1 & 15.15(.02) & 14.46(.02) & 14.47(.03) & 14.51(.03) \\
6474.9 & 27.1 & 15.24(.03) & 14.47(.03) & 14.54(.03) & 14.69(.04) \\
6475.9 & 28.1 & 15.41(.03) & 14.57(.02) & 14.60(.03) & 14.66(.04) \\
6477.9 & 30.1 & 15.61(.03) & 14.71(.02) & 14.70(.03) & 14.73(.03) \\
6482.9 & 35.1 & 16.20(.06) & 15.19(.05) & 15.06(.06) & 14.99(.07) \\
6483.9 & 36.1 & 16.42(.06) & 15.22(.04) & 15.05(.05) & 15.06(.06) \\
6484.9 & 37.1 & 16.26(.06) & 15.20(.04) & 15.22(.06) & 15.21(.07) \\
6487.9 & 40.1 & 16.60(.11) & 15.29(.08) & 15.20(.08) & 15.03(.09) \\
6489.9 & 42.1 & 16.37(.10) & 15.42(.07) & 15.29(.08) & 15.16(.09) \\
6490.9 & 43.1 & 16.77(.14) & 15.47(.11) & 15.61(.13) & 15.46(.15) \\
6491.9 & 44.1 & 16.81(.21) & 15.43(.16) & 15.30(.17) & 15.15(.18) \\
\hline
\end{tabular}
\end{center}
\smallskip
{\bf Notes.} $^{(a)}$JD$-$2,450,000. $^{(b)}$ With respect to $t_{0}$ = 2,456,447.8 $\pm$ 0.5 JD. Errors are given in parentheses.
\end{table*}

\begin{table*}
\begin{center}
\caption{Unfiltered ROTSE-IIIb magnitudes of SN2013df}
\label{tab:rotse_lc}
\begin{tabular}{ccc}
\hline
\hline
MJD $-$ 50\,000 & Phase$^a$ (days) & Magnitude \\
\hline
6459.2 & 11.9 & 14.88(.26) \\
6460.2 & 12.9 & 14.43(.36) \\
6461.3 & 14.0 & 14.51(.35) \\
6470.3 & 23.0 & 14.27(.49) \\
6473.2 & 25.9 & 14.01(.25) \\
6478.2 & 30.9 & 14.44(.13) \\
6479.2 & 31.9 & 14.66(.37) \\
6481.2 & 33.9 & 14.80(.17) \\
6482.2 & 34.9 & 14.66(.34) \\
6483.2 & 35.9 & 14.95(.07) \\
6484.2 & 36.9 & 14.98(.12) \\
6484.2 & 36.9 & 15.10(.11) \\
6485.2 & 37.9 & 15.46(.12) \\
6486.2 & 38.9 & 15.04(.21) \\
6487.2 & 39.9 & 15.54(.50) \\
6488.2 & 40.9 & 15.28(.16) \\
6497.2 & 49.9 & 15.45(.35) \\
6505.2 & 57.9 & 15.67(.10) \\
6506.2 & 58.9 & 15.37(.25) \\
6507.2 & 59.9 & 15.47(.20) \\
6508.2 & 60.9 & 15.72(.28) \\
6512.2 & 64.9 & 15.72(.17) \\
6514.2 & 66.9 & 15.37(.20) \\
6516.2 & 68.9 & 15.67(.11) \\
6518.2 & 70.9 & 16.24(.46) \\
6530.1 & 82.8 & 16.72(.71) \\
6615.5 & 168.2 & 18.01(.31) \\
\hline 
\hline
\end{tabular}
\end{center}
\smallskip
{\bf Notes.} $^{(a)}$ With respect to $t_{0}$ = 2,456,447.8 $\pm$ 0.5 JD. Errors are given in parentheses.
\end{table*}

\label{lastpage}

\end{document}